\documentclass[a4paper]{article}
\usepackage[top=30truemm,bottom=30truemm,left=25truemm,right=25truemm]{geometry}

\usepackage{indentfirst}
\usepackage[pdftex]{graphicx}
\usepackage[hang,small,bf]{caption}
\usepackage[subrefformat=empty]{subcaption}
\usepackage[italicdiff]{physics}

\usepackage[compat=1.1.0]{tikz-feynhand}

\usepackage{amsmath,amssymb}
\usepackage{bm,color}
\usepackage{slashed}
\usepackage{cite}

\usepackage[colorlinks=true,allcolors=blue]{hyperref}

\hypersetup{%
setpagesize=false,
 bookmarksnumbered=true,%
 bookmarksopen=true,%
 colorlinks=true,%
 linkcolor=blue,
 citecolor=red,
}

\begin{document}
\titlepage
\today
  \begin{flushright}
   {\bf
  \begin{tabular}{l}
OCHA-PP-375
  \end{tabular}
   }
  \end{flushright}

\vspace*{1.0cm}
\baselineskip 18pt
\begin{center}
 {\Large \bf
Analyzing cancellation mechanism of the dark matter-quark scattering in a complex singlet extension of the Standard Model
}

\vspace*{1.0cm}

 {\bf
 Gi-Chol Cho$^a$ and Chikako Idegawa$^b$
}

\vspace*{0.5cm}
$^a${\em Department of Physics, Ochanomizu University, Tokyo 112-8610,
 Japan}
 \\
$^b${\em Graduate school of Humanities and Sciences, Ochanomizu
 University, Tokyo 112-8610, Japan}
 \\
\end{center}

\vspace*{1cm}

\baselineskip 18pt
\begin{abstract}
 \noindent
We investigate a suppression mechanism of dark matter and quark scattering amplitudes in a complex singlet extension of the Standard Model. 
It has been pointed out that, in a some variant of the model, 
the scattering amplitudes cancel each other in the limit in which two mediator scalars degenerate in their masses. 
We study the origin of such the cancellation mechanism and show that the operators describing the Higgs-singlet scalar mixing play essential role. 
We derive sum rules for couplings in the general scalar potential of the model, which guarantee the cancellation of the scattering amplitudes in the tree and the 1-loop level. 
\end{abstract}

\newpage
\section{Introduction}

The existence of dark matter (DM) provides direct evidence for new physics beyond the Standard Model (SM). One of the leading candidates for DM is the weakly interacting massive particle (WIMP), which is known to be in good agreement with data on the thermal abundance of dark matter obtained from observations of the cosmic microwave background radiation (so-called ``WIMP miracle''). However, while astronomical observations have confirmed the existence of DM, no WIMP-DM signal has been obtained either at high-energy accelerator experiments such as the LHC and/or at so-called direct detection experiments. In particular, the recent LUX-ZEPLIN (LZ) experiment~\cite{LZ:2022ufs} provides strict upper bounds on the spin-independent cross section between DM and nucleons. Therefore, an important task in the construction of DM models is to suppress the scattering of DM and nucleons.
Several ideas have been proposed to naturally suppress the scattering of DM and nucleons~\cite{Ipek:2014gua, Escudero:2016gzx, Abe:2018emu, Gross:2017dan, Chao:2017vrq, Ghorbani:2018pjh, Jiang:2019soj,  Abe:2021nih, Cai:2021evx, Abe:2022mlc, Maleki:2022zuw, Liu:2022evb}, including the so-called pseudo Nambu-Goldstone (pNG) DM model~\cite{Gross:2017dan, Jiang:2019soj, Abe:2021nih, Liu:2022evb}. 
This model extends the SM by introducing a complex singlet scalar $S$. The imaginary part of $S$ becomes a pNG boson due to the global U(1) symmetry breaking of the scalar potential. The CP-symmetry of the scalar potential forbids a decay of the pNG boson so that it plays the role of DM. 
On the other hand, a real part of $S$ mixes with the SM Higgs boson, and there are two CP-even mass eigenstates $h_1$ and $h_2$. Then the DM-quark scattering processes are given by two scattering amplitudes, one mediated by $h_1$ and the other by $h_2$. 
The authors of ref.~\cite{Gross:2017dan} proposed the simplest version of this model (hereafter, we call it the minimal pNG-DM model) where the global U(1) symmetry of the scalar potential is softly broken by a dimension-2 operator $S^2$. 
Then, the scattering amplitudes mediated by $h_1$ and $h_2$ cancel independently of the DM coupling and mass in 
the limit of zero momentum transfer. 
However, since the scalar potential of the minimal pNG-DM model has a discrete symmetry $S \to -S$, the model is suffered from the so-called domain-wall problem when $S$ develops the vacuum expectation value (VEV). To avoid this problem, the authors of ref.~\cite{Abe:2021nih} proposed introducing a linear term of $S$ to the scalar potential. As a result, it is shown that the suppression mechanism in ref.~\cite{Gross:2017dan} no longer works, and the cancellation of the amplitudes could be achieved when masses of two mediator particles ($h_1$, $h_2$) degenerate. This is called the degenerate scalar scenario. In the scenario, the interactions of $h_1$ ($h_2$) with SM particles are given by the SM Higgs couplings multiplied by $\cos{\alpha}$ ($- \sin{\alpha}$) where $\alpha$ is the mixing angle. When the masses of the two are close, it is hard to experimentally distinguish between $h_1$ and $h_2$. The possibility of verifying such a degenerate scalar scenario in future collider experiments~\cite{Abe:2021nih}, strong first-order electroweak phase transitions~\cite{Cho:2021itv}, observation of the gravitational wave~\cite{Cho:2022our} have been investigated. Also, the realization of the degenerate scalar scenario has been analyzed based on the multi-critical point principle~\cite{Froggatt:1995rt,Kannike:2020qtw} in ref.~\cite{Cho:2022zfg}. 
In the SM, the FCNC processes vanish in the degenerate mass limit of quarks, mediating the processes in the 1-loop diagrams. This is called the GIM mechanism, and it is known that the origin of the GIM mechanism is the unitarity of the CKM matrix, which mixes the different quark flavors in the charged current interactions. 
In one aspect, the degenerate scalar scenario seems similar to the GIM mechanism since the cancellation of DM scattering amplitudes is achieved in the degenerate mass limit of two mediators $(h_1,h_2)$, due to the orthogonality of the mixing matrix of the SM Higgs and the singlet scalar. 
However, the charged current interactions of quarks mediated by the $W$-boson are controlled by the gauge symmetry, whereas the interactions in the scalar potential of the CxSM could be some variants. 
Therefore, it is worth systematically examining the origin of the cancellation mechanism in the degenerate scalar scenario. 
In this paper, we investigate the origin of the cancellation mechanism of DM-quark scattering amplitudes in the degenerate scalar scenario of the CxSM. Our study is based on a model-independent framework for the DM-quark scattering processes. It is found that some combinations of couplings in the scalar potential are required to satisfy a certain condition to make the amplitudes cancel in the degenerate mass limit of two scalar mediators. The condition tells us that operators representing the mixing between the SM Higgs and the singlet scalar are constrained to cancel the amplitudes. We also study the cancellation mechanism in the degenerate scalar scenario in the 1-loop level. 
The sum rules for cancellation of the 1-loop amplitude in the degenerate scalar limit will be shown. 
We show that the most general CxSM is incompatible with the cancellation mechanism. On the other hand, the pNG-DM model satisfies the relationships of the couplings to cancel the DM-quark amplitude at the tree-level and in the 1-loop level. 
The 1-loop corrections for the DM-quark scattering in the minimal pNG-DM model are worth mentioning for comparison. 
It has been pointed out that the 1-loop corrections violate the cancellation of amplitudes~\cite{Azevedo:2018exj, Ishiwata:2018sdi, Alanne:2020jwx, Glaus:2020ihj, Abe:2022mlc}. However, the scattering cross sections presented in the literatures vanishes when the masses of $h_1$ and $h_2$ degenerate, which is consistent with our results. 
This paper is organized as follows. In Sec.~\ref{sec;model}, we introduce the general CxSM and explain the relationship between the coefficients of interactions. In Sec.~\ref{sec;tree}, we describe the tree-level suppression mechanism of DM-quark scattering and the conditions required for scalar potentials. Sec.~\ref{sec;loop} further introduces cancellation in the degenerate scalar scenario at the 1-loop level using the pNG-DM model. Sec.~\ref{sec;sum} is devoted to summarizing our study.

\section{Scalar potential in the CxSM}\label{sec;model}
In this section, we briefly review the scalar potential in the CxSM to fix our notation. 
The CxSM consists of the SM particles and a complex singlet scalar field. 
The most general scalar potential is given by~\cite{Barger:2008jx}
\begin{align}
 V(H,S) 
 &=
 \frac{m^2}{2}H^\dagger H + \frac{\lambda}{4}\qty(H^\dagger H)^2 
+ 
\frac{\delta_2}{2}H^\dagger H |S|^2
+\frac{b_2}{2}|S|^2
+ \frac{d_2}{4}|S|^4 
\nonumber \\
&+ \left(
|a_1| e^{i\phi_{a_1}} S 
 + \frac{|b_1| e^{i\phi_{b_1}}}{4} S^2 
+ 
\frac{|\delta_3 |e^{i\phi_{\delta_3}}}{4}H^\dagger H S^2 
 +
\frac{|\delta_1|e^{i\phi_{\delta_1}}}{4}H^\dagger H S 
+
 \frac{|c_1| e^{i\phi_{c_1}}}{6} S^3 
\right.
\nonumber \\
&
\left.
+
 \frac{|c_2| e^{i\phi_{c_2}}}{6} S|S|^2 
 +
 \frac{|d_1| e^{i\phi_{d_1}}}{8} S^4 
 +
 \frac{|d_3| e^{i\phi_{d_3}}}{8} S^2|S|^2 + \mathrm{c.c.}
\right ), 
\label{gnrlp}
\end{align}
where $H$ and $S$ denote the SM Higgs and a singlet scalar fields, respectively. 
The operators in the first line of r.h.s. in \eqref{gnrlp} are invariant under a global U(1) transformation of $S$, and those in the second and third lines break the symmetry. 
Two scalar fields, $H$ and $S$, are parametrized as 
\begin{align}
 H = \frac{1}{\sqrt{2}}
\qty(
\begin{matrix}
 0\\
v+h
\end{matrix}
), 
~~~
S = \frac{1}{\sqrt{2}}\qty(v_S + s + i\chi),
\end{align}
where $v$ and $v_S$ are vacuum expectation values (VEVs), $h$ is the Higgs boson that would mix with a singlet scalar $s$. 
Hereafter we assume that the scalar potential is invariant under the CP transformation $S\to S^{*}$. 
Then, all phases in \eqref{gnrlp} are zero ($\phi_{a_1}=\phi_{b_1}=\phi_{\delta_3}=\phi_{\delta_1}=\phi_{c_1}=\phi_{c_2}=\phi_{d_1}=\phi_{d_3}=0$). 
As a consequence of the CP invariance of the scalar potential, the pseudoscalar $\chi$ becomes stable and behaves as a DM. 
Under the CP invariance of the scalar potential, two CP-even scalars $h$ and $s$ mix through VEVs. We define the mass eigenstates $(h_1,h_2)$ by 
\begin{align}
 \begin{pmatrix}
  h_1\\
  h_2
 \end{pmatrix}
=
\begin{pmatrix}
 \cos{\alpha} & \sin{\alpha}
\\
 -\sin{\alpha} & \cos{\alpha}
\end{pmatrix}
\begin{pmatrix}
 h\\
 s
\end{pmatrix}
, 
\label{trns}  
\end{align}
where $\alpha$ being the mixing angle, and we define $h_1$ as the scalar boson observed at the LHC with mass $m_{h_1}=125~\mathrm{GeV}$. Although we assume the CP invariance of the scalar potential, the CP-violating sources in the full Lagrangian may break the CP symmetry of the scalar potential through the radiative corrections, and the stability of DM is not guaranteed. However, this issue is beyond the scope of this paper.
It is useful for our study to express the scalar potential $V$ with the CP-symmetry in terms of the component fields as 
\begin{align}
 V &= \frac{1}{2}C_{ij} \phi_i \phi_j 
+ C_{ijk} \phi_i \phi_j \phi_k
+ C_{ijkl} \phi_i \phi_j \phi_k \phi_l, 
\label{spcoef}
\end{align}
where for the couplings of current eigenstates, the index $i$ of the coefficient denotes any one of $(h,s,\chi)$, and $\phi_h=h$, etc., while as for the couplings of mass eigenstates, the index $i$ indicates $h_1,h_2$ and $\chi$. 
Note that, for simplicity, we sometimes use the abbreviation such as $C_{111} \equiv C_{h_1 h_1 h_1}$, etc. 
The coefficients $C_{ij}, C_{ijk}$ and $C_{ijkl}$ can be expressed in terms of coefficients of operators in the scalar potential \eqref{gnrlp} with the CP-invariance. 
Using the coefficients of bilinear terms $C_{ij}$, the mass terms of two CP-even scalars $(h,s)$ can be expressed as 
\begin{align}
 -\mathcal{L}
 &\supset 
\frac{1}{2}\qty(
 C_{hh} h^2 + C_{ss} s^2 + 2C_{hs} h s
)
\nonumber \\
&=
\frac{1}{2}\qty(
 C_{1 1} h_1^2 +  C_{2 2} h_2^2
), 
\label{massmatrix}
\end{align}
where $C_{1 1}$ and $C_{2 2}$ are identified with the mass eigenvalues $m_{h_1}^2$ and 
$m_{h_2}^2$, respectively: 
\begin{align}
C_{ii} &= m_{h_i}^2~~~(i=1,2), 
\\
m_{h_{1,2}}^2
&=
 \frac{1}{2}\qty{
C_{hh} + C_{ss} \mp \sqrt{
 \qty(C_{hh}+C_{ss})^2 - 4\qty(C_{hh} C_{ss} - C_{hs}^2)
 }
 }. 
\end{align}
For later convenience, we summarize relations of coefficients $C_{ij}$ between the current and mass eigenstates:  
\begin{align}
 C_{hh} \cos{\alpha}  + C_{hs} \sin{\alpha} 
 &=
 C_{1 1} \cos{\alpha} ,
\\
 -C_{hh} \sin{\alpha}  + C_{hs} \cos{\alpha} 
 &=
 -C_{2 2} \sin{\alpha} ,
\\
 C_{hs} \cos{\alpha}  + C_{ss} \sin{\alpha} 
 &=
 C_{1 1} \sin{\alpha} ,
\label{chs1}
\\
 -C_{hs} \sin{\alpha}  + C_{ss} \cos{\alpha} 
 &=
 C_{2 2} \cos{\alpha} .
\label{chs2}
\end{align}
We collect the interactions relevant to the DM-quark scattering amplitudes. 
The Yukawa interactions of a fermion $f$ and $h_1$ or $h_2$ are given by
\begin{align}
 -\mathcal{L}_{\mathrm{Yukawa}}
&=
 C_{ff h_1} \overline{f} f h_1 + C_{ff h_2} \overline{f} f h_2
  + \mathrm{h.c.},  
\end{align}
where  
\begin{align}
 C_{ff h_1} &= \frac{\sqrt{2} m_f}{v} \cos{\alpha}, 
~~~
 C_{ff h_2} = -\frac{\sqrt{2} m_f}{v} \sin{\alpha},
\label{h1h2yukawa}
\end{align}
and $m_f$ represents a mass of fermion $f$. 
The scalar trilinear interactions between $\chi$ and $h_1$ ($h_2$) are restricted by $\chi\chi h_1$ and $\chi\chi h_2$ due to the CP invariance of the scalar potential. The interaction is described by 
\begin{align}
-\mathcal{L}
&\supset 
 C_{\chi\chi h} h \chi^2 +  C_{\chi\chi s} s \chi^2
\nonumber \\
&=
C_{\chi\chi h_1} h_1 \chi^2 + C_{\chi\chi h_2} h_2 \chi^2,
\end{align}
where 
\begin{align}
C_{\chi\chi h_1} &\equiv C_{\chi\chi h} \cos \alpha + C_{\chi\chi s} \sin{\alpha}, 
~~~
C_{\chi\chi h_2} \equiv -C_{\chi\chi h} \sin \alpha + C_{\chi\chi s} \cos{\alpha}. 
\label{ccssdd}
\end{align}
The pNG-DM model, which has been adopted to demonstrate the degenerate scalar scenario~\cite{Abe:2021nih}, requires coefficients of some operators in the general scalar potential \eqref{gnrlp} to be zero: 
\begin{align}
 V_0 (H,S)
&=
\frac{m^2}{2}H^\dagger H + \frac{\lambda}{4} \qty(H^\dagger H )^2 
+ \frac{\delta_2}{2}H^\dagger H |S|^2 
+ \frac{b_2}{2}|S|^2
+\frac{d_2}{4}|S|^4
+ \qty(
a_1 S + \frac{b_1}{4}S^2 + \mathrm{c.c.}
).
\label{mnmlmdl}
\end{align}
Both $a_1$ and $b_1$ softly break the global U(1) symmetry of $S$, and the former is introduced to prevent the domain-wall problem. When the scalar potential has CP symmetry, i.e., $a_1$ and $b_1$ are real, the pseudoscalar $\chi$ is a DM candidate. 
In the minimal pNG-DM model~\cite{Gross:2017dan}, only $b_1$ is allowed to be non-zero, i.e. $a_1=0$ in \eqref{mnmlmdl}, so that there is a discrete symmetry $S \to -S$ which is broken spontaneously by the VEV $v_S$, and is suffered from the domain-wall problem. 
\section{Tree-level condition for suppression of the DM-quark scattering}
\label{sec;tree}
\begin{figure}[htpb]
\centering
\begin{tikzpicture}
\begin{feynhand}
\vertex (a) at (0,0); 
\vertex (b) at (-1,0.8); 
\vertex (beta) at (-1,1.6); 
\vertex (c) at (-1,2.5); 
\vertex[particle] (d) at (-2,1.6){$\chi$}; 
\vertex[particle] (e) at (2,1.6){$\chi$}; 
\vertex (f) at (0,-2); 
\vertex[particle] (g) at (-2,-3.6){$q$};
\vertex[particle] (h) at (2,-3.6){$q$};
\propag[sca] (a) to (d);\propag[sca] (a) to (e);
\propag[sca] (a) to [edge label=$h_{1,2}$] (f);
\propag[fermion] (g) to (f); \propag[fermion] (f) to (h);
\propag[fermion, mom={$p_1$}] (g) to (f);
\propag[fermion, mom={$p_3$}] (f) to (h);
\propag[scalar, mom'={$p_2$}] (d) to (a);
\propag[scalar, mom'={$p_4$}] (a) to (e);
\end{feynhand}
\end{tikzpicture}
\caption{Feynman diagram of the scattering process $\chi q \to \chi q$ mediated by $h_1$ and $h_2$.}
\label{fig;tree}
\end{figure}
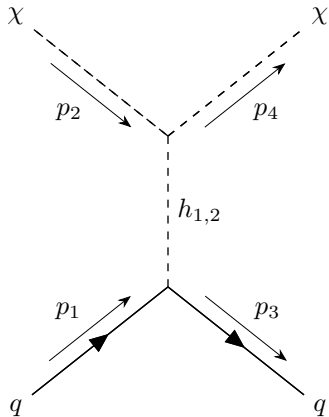
\begin{figure}[htpb]
\centering
\includegraphics[width=15cm]{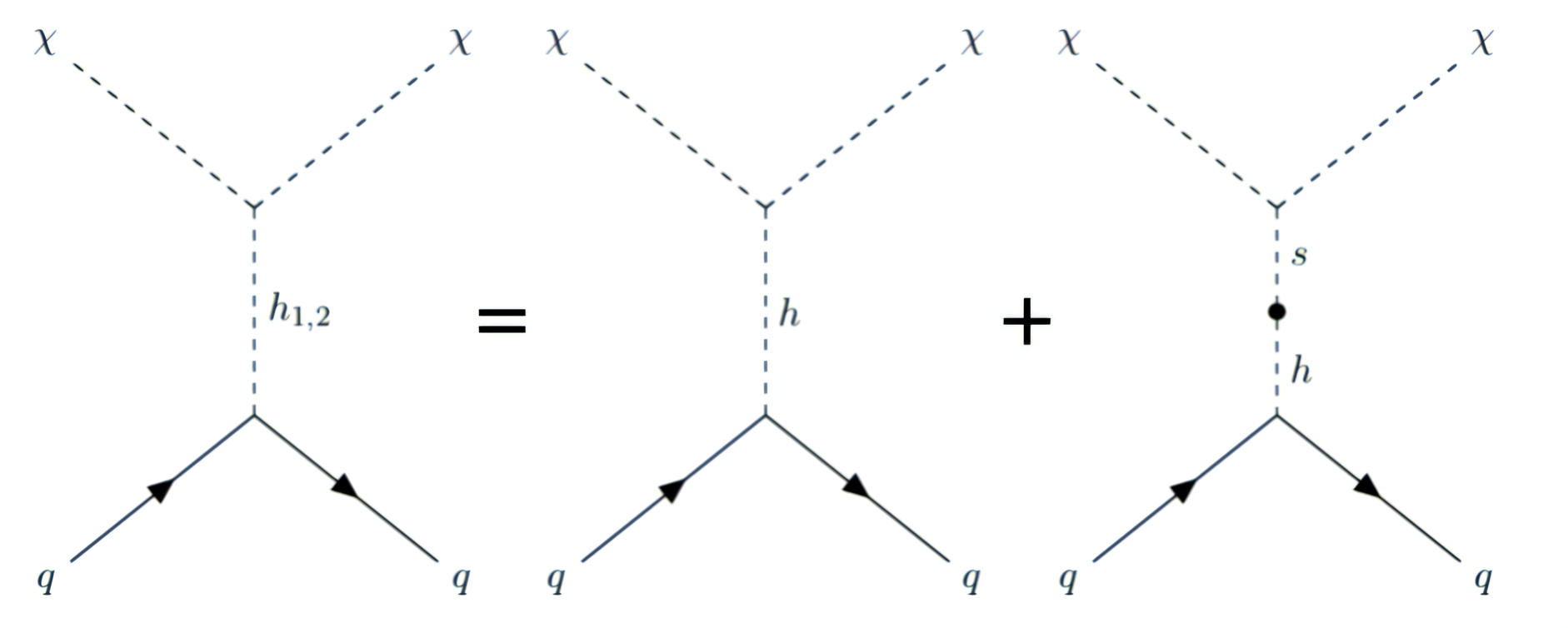}
\caption{Feynman diagrams of the DM-quark scattering in the current eigenstates of scalar mediators.}
\label{fig;trcur}
\end{figure}

In this section, we study a suppression mechanism of the scattering process of DM $\chi$ off a quark $q$ in the CxSM, described by the Feynman diagram in Fig.~\ref{fig;tree}. 
The scattering amplitude $\mathcal{M}$ is given by a sum of two amplitudes $\mathcal{M}_1$ and $\mathcal{M}_2$
mediated by $h_1$ and $h_2$, respectively: 
\begin{align}
 i\mathcal{M} 
&=
i\qty(\mathcal{M}_1 + \mathcal{M}_2), 
\label{msumtr}
\\
 i\mathcal{M}_1 
&= 
-i 2 C_{\chi\chi h_1} C_{qq h_1}  \frac{1}{t-m_{h_1}^2} \bar{u}(p_3) u(p_1), 
\label{m1tr}
\\
 i\mathcal{M}_2 
&= 
-i 2 C_{\chi\chi h_2} C_{qq h_2}  \frac{1}{t-m_{h_2}^2} \bar{u}(p_3) u(p_1), 
\label{m2tr}
\end{align}
where $t$ describes a momentum transfer, $t\equiv \qty(p_1 - p_3)^2$, and $u(p)~(\bar{u}(p))$ represents an incoming (outgoing) quark spinor with a momentum $p$. 
A factor 2 in the r.h.s of \eqref{m1tr} and \eqref{m2tr} is a symmetry factor for the $\chi\chi h_i$ vertex. 
Since the momentum transfer $t$ in the direct detection experiments is very small as compared to the mediator masses, i.e., $t \ll m_{h_1}^2, m_{h_2}^2$, 
the amplitude \eqref{msumtr} can be written by 
\begin{align}
 i\mathcal{M} 
&=
2i \bar{u}(p_3) u(p_1) 
\frac{\sqrt{2}m_q}{v}
\qty(
\frac{1}{m_{h_1}^2}
 C_{\chi\chi h_1} \cos \alpha - 
\frac{1}{m_{h_2}^2}
C_{\chi\chi h_2} \sin{\alpha}
), 
\label{cndtn}
\end{align}
where the Yukawa couplings \eqref{h1h2yukawa} are used instead of $C_{qqh_1}$ and $C_{qqh_2}$, and 
note that $\qty(C_{\chi\chi h_1}, C_{\chi\chi h_2})$ can be replaced by $\qty(C_{\chi\chi h}, C_{\chi\chi s})$ as given in \eqref{ccssdd}. 
The general scalar potential in the CxSM~\eqref{gnrlp} allows us to rewrite the trilinear couplings $C_{\chi\chi h}$ and $C_{\chi\chi s}$ by the bilinear couplings $C_{hs}$ and $C_{ss}$ as 
\begin{align}
 C_{\chi\chi h} &= \frac{A}{v_s} \qty(C_{hs} + \Delta_h), 
~~~
 C_{\chi\chi s} =
  \frac{A}{v_s} \qty(C_{ss} + \Delta_s), 
\label{aaii}
\end{align}
where parameters $A, \Delta_h$ and $\Delta_s$ are given by parameters besides $C_{hs}$ and $C_{ss}$ in the scalar potential. 
Taking account of Eqs.~\eqref{chs1}, \eqref{chs2} and \eqref{aaii}, we find that the inside of parentheses in \eqref{cndtn} can be written as 
\begin{align}
 \frac{C_{h_1 \chi\chi} }{m_{h_1}^2} \cos{\alpha}
-
 \frac{C_{h_2 \chi\chi} }{m_{h_2}^2} \sin{\alpha}
&=
 \frac{A}{v_s}
\qty[
\sin{\alpha} \cos{\alpha}
 \qty(
 \frac{m_{h_1}^2 + \Delta_s}{m_{h_1}^2}
-
 \frac{m_{h_2}^2 + \Delta_s}{m_{h_2}^2}
 )
+ 
 \Delta_h 
\qty(
\frac{\cos^2\alpha}{m_{h_1}^2}+\frac{\sin^2\alpha}{m_{h_2}^2}
)
]. 
\end{align}
Therefore, the amplitude \eqref{cndtn} vanishes in the degenerate scalar scenario $(m_{h_1}=m_{h_2})$ when 
$\Delta_h$ satisfies 
\begin{align}
\Delta_h=0. 
\label{cnclcnd}
\end{align}
In other words, the DM-quark scattering amplitude is not suppressed in the general CxSM at the degenerate limit of mediator masses, unlike the pNG-DM model. 
The suppression of the DM-quark scattering requires the coupling $C_{\chi\chi h}$ should be given by 
$C_{hs}$ only (see, \eqref{aaii} and \eqref{cnclcnd}). 
It is easy to understand the origin of the condition \eqref{cnclcnd} 
using the Feynman diagrams in the current eigenstates. 
In Fig.~\ref{fig;trcur}, we depict the Feynman diagrams of the DM-quark scattering in the current eigenstates of mediators. 
Since the singlet scalar $s$ cannot couple to quark $q$, the scalar $s$ emitted from the DM $\chi$ should be converted to the SM Higgs $h$ through the coupling $C_{hs}$ to propagate the DM-quark scattering (see, the right diagram in Fig.~\ref{fig;trcur}) . On the other hand, the amplitude mediated by the SM Higgs $h$ is proportional to the coupling $C_{\chi\chi h}$ (the middle diagram in Fig.~\ref{fig;trcur}). 
Therefore, two diagrams can be canceled only when $C_{hs}$ and $C_{\chi\chi h}$ have a specific relation. 
Such a relationship is possible when the origin of the bilinear $\qty(C_{hs})$ and the trilinear $\qty(C_{\chi\chi h})$ couplings is common, and it is why the condition \eqref{cnclcnd} is required. 
For example, we show that the tree-level condition \eqref{cnclcnd} is satisfied in the pNG-DM model \eqref{mnmlmdl}. 
In this model, couplings $C_{hs}$ and $C_{\chi\chi h}$, which represent the 
the mixing between the SM Higgs $h$ and the singlet $s$ (or the DM $\chi$) 
are given by 
 \begin{align}
 V_0 (H,S) 
 &\supset
 \frac{\delta_2}{2}\qty|H|^2 \qty|S|^2
\nonumber \\
 &\supset 
C_{\chi\chi h} h \chi^2 + C_{hs} hs, 
\end{align}
where two couplings are given by
\begin{align}
C_{\chi\chi h} &= \frac{\delta_2}{4}v,
\label{c_cch}
\\
C_{hs} &= \frac{\delta_2}{2} v_s v. 
\label{c_hs}
\end{align}
Comparing \eqref{c_cch} and \eqref{c_hs} with \eqref{aaii}, we find 
\begin{align}
A &=\frac{1}{2}, 
~~~
\Delta_h = 0. 
\label{cndtnmnml}
\end{align}
Although \eqref{cndtnmnml} is enough to show that the DM-quark scattering in the pNG-DM model vanishes when $m_{h_1}=m_{h_2}$, we also evaluate another parameter $\Delta_s$ \eqref{aaii} 
for completeness.  
The couplings $C_{\chi \chi s}$ and $C_{ss}$ are given as follows
\begin{align}
 V_0 (H,S)
 &\supset 
\frac{\delta_2}{2}\qty|H|^2 \qty|S|^2 + \frac{b_2}{2}\qty|S|^2 
+  \frac{d_2}{4} \qty|S|^4 + \frac{b_1}{4} \qty( S^2 + S^{*2})
\nonumber \\
&=
C_{\chi\chi s} s\chi^2 + \frac{1}{2}C_{ss} s^2, 
\end{align}
where $C_{\chi\chi s}$ and $C_{ss}$ are given by 
\begin{align}
C_{\chi\chi s}  
 &= 
 \frac{d_2}{4} v_s, 
\\
C_{ss} 
 &=
\frac{3}{4}d_2 v_S^2 + \frac{\delta_2}{4} v^2  + \frac{b_1}{2} + \frac{b_2}{2}. 
\end{align}
Since \eqref{cndtnmnml} requires $A=\frac{1}{2}$, $\Delta_s$ is given as 
\begin{align}
 \Delta_s 
 &= -\frac{d_2}{4} v_s^2 - \frac{\delta_2}{4}v^2 - \frac{b_1+b_2}{2} 
\nonumber \\
 &= \sqrt{2} \frac{a_1}{v_S}, 
\label{dels}
\end{align}
where we used a minimization condition of the scalar potential $\qty(\pdv{V_0}{S}=0)$ which gives 
the following identity: 
\begin{align}
-b_2= \frac{\delta_2}{2}v^2 + \frac{d_2}{2}v_S^2 + b_1 + 2\sqrt{2} \frac{a_1}{v_S}. 
\end{align}
It should be noted that in the minimal pNG-DM model, i.e., $a_1=0$ in \eqref{mnmlmdl}, 
the parameter $\Delta_s$~\eqref{dels} is zero and \eqref{aaii} tells us that the DM-quark scattering amplitude vanishes even for $m_{h_1} \neq m_{h_2}$. 
Next, we extend the above analysis to the general CxSM whose scalar potential  contains the self-interaction terms of $S$ and various $H$-$S$ mixing terms in addition to the terms in the scalar potential of the minimal pNG-DM model. 
We learn from the condition \eqref{cnclcnd} that $\Delta_s$ in \eqref{aaii} is irrelevant to the cancellation of the scattering amplitudes. 
Thus, the trilinear and quartic terms of $S$ in the general scalar potential \eqref{gnrlp} are harmless since those terms only contribute to $\Delta_s$. 
However, the condition \eqref{cnclcnd} is sensitive to the $H$-$S$ mixing terms. 
The couplings $C_{\chi\chi h}$ and $C_{hs}$ in the scalar potential of the general CxSM \eqref{gnrlp} are given by 
\begin{align}
 V (H,S)  
 &\supset
 \frac{\delta_2}{2}\qty|H|^2 \qty|S|^2 
+
\qty(
 \frac{\delta_1}{4} \qty|H|^2 S
+ \frac{\delta_3}{4} \qty|H|^2 S^2
+ \mathrm{c.c.}
)
\nonumber \\
&=
C_{\chi\chi h} h\chi^2 + C_{hs} hs, 
\end{align} 
where
\begin{align}
 C_{\chi\chi h} 
 &= 
 \frac{\delta_2-\delta_3}{4}v, 
\label{gcch}
\\
 C_{hs} 
 &= 
 \frac{v}{2} \qty(
\frac{\delta_1}{\sqrt{2}} + \delta_2 v_S + \delta_3 v_S
). 
\label{ghs}
\end{align}
Since $\delta_1$ appears only in $C_{hs}$ \eqref{ghs}, we emphasize that non-zero $\delta_1$ conflicts with the proportionality between Eqs.~\eqref{gcch} and \eqref{ghs} which is required by Eqs.\eqref{aaii} and \eqref{cnclcnd}. 
Comparing \eqref{gcch} with \eqref{ghs}, 
we find the condition $\Delta_h = 0$~\eqref{cnclcnd} requires one of the following conditions
\begin{align}
\mathrm{(i)}~~
\delta_1&=\delta_2=0, 
\\
\mathrm{(ii)}~~
\delta_1 &=\delta_3=0.
\label{cond3}
\end{align}
Note that $\delta_2$ preserves the global U(1) symmetry while $\delta_1$ and $\delta_3$ break it softly. Condition (ii) corresponds to the pNG-DM model. 
From the symmetry point of view, condition (i) is unnatural because requiring $\delta_2=0$ does not recover any symmetry. 
For completeness, we give couplings $C_{\chi\chi s}$ and $C_{ss}$
\begin{align}
 C_{\chi\chi s}
&=
\frac{1}{2\sqrt{2}}\qty(-c_1 + \frac{c_2}{3}) 
+ 
\frac{d_2 - 3d_1}{4}v_S, 
\\
 C_{ss}
&=
\frac{b_1+b_2}{2}
+
\frac{v_S}{\sqrt{2}}\qty(c_1 + c_2)
+
\frac{3}{4}\qty(d_1 + d_2 + d_3) v_S^2
+
\frac{\delta_2+\delta_3}{4}v^2
. 
\end{align}

In this section, we investigated the condition to hold the cancellation mechanism of the DM-quark scattering processes in the CxSM with degenerate scalars. It is found that the operators in the scalar potential describing the $H$-$S$ mixing play an important role in the cancellation. 
Although there are three operators for the $H$-$S$ mixing in the general CxSM whose couplings are given by $\delta_1,~\delta_2$ and $\delta_3$, we found that the cancellation mechanism works when coupling either $\delta_2$ or $\delta_3$ is present, and co-existence of both couplings are not allowed for the cancellation. 
On the other hand, various operators describing the self-interactions of the singlet $S$ are irrelevant to the cancellation. Therefore, the pNG-DM model~\eqref{mnmlmdl} with degenerate scalars is a successful and minimal model to realize the cancellation mechanism.

\section{Cancellation of DM-quark scattering in the degenerate scalar scenario beyond the tree-level}\label{sec;loop}
In this section, we study the 1-loop contributions to the DM-quark scattering in the degenerate scalar scenario. 
It was pointed out in the previous section that some of the operators in the scalar potential of the general CxSM should be forbidden for the cancellation of amplitudes at the tree-level, so we proceed our study in this section based on the pNG-DM model \eqref{mnmlmdl} which is known to satisfy the condition of the cancellation mechanism. 
This section aims to examine whether the 1-loop diagrams provide finite contributions to the scattering processes in the degenerate scalar limit, $m_{h_1}=m_{h_2}$.\footnote{Since the tree-level amplitudes vanish at the limit, no divergence appears at the loop-level, and the result should be finite. }
Since, in the pNG-DM model, the coupling $\delta_2$ in \eqref{mnmlmdl} is only source of the $H$-$S$ mixing, it is useful to give $\delta_2$ in terms of mass eigenvalues for later discussion: 
\begin{align}
 \delta_2 &= \frac{2}{v v_S}\qty(m_{h_1}^2-m_{h_2}^2) \sin\alpha \cos\alpha. 
\label{dlt2}
\end{align} 
Thus $\delta_2$ is zero in the degenerate scalar limit. 
We classify the 1-loop diagrams for the DM-quark scattering into five groups as summarized in Figs.~\ref{type1}-\ref{type5}. 
In most diagrams, the loops consist of the DM $\chi$ and/or the scalars $h_1,~h_2$. 
However, the SM particles could appear in the loop diagrams in Figs.~\ref{type1}-\ref{type3} (denoted by $A$). Note that Figs.~\ref{type1} and \ref{type5} include diagrams for the DM-gluon scattering at the 2-loop level. Although these are sub-dominant contributions, we added these diagrams for comparison with the results in Ref.~\cite{Ishiwata:2018sdi,Azevedo:2018exj}. 
In the following subsections, we discuss the structural characteristics of the diagrams and show that the diagrams belonging to each group disappear in the limit $m_{h_1}=m_{h_2}$. 
For our purpose, the amplitudes given in the following subsections are shown in the non-relativistic limit of the momentum transfer $t\to 0$ and the degenerate scalar limit $m_{h_1}=m_{h_2} \equiv m$. 

\subsection{Group-1}
\begin{figure}[htpb]
\begin{minipage}[b]{0.2\linewidth}
\centering
\begin{tikzpicture}[scale=0.8]
\begin{feynhand}
\vertex (a) at (0,0);
\vertex (b) at (0,-0.5);
\vertex (c) at (0,-1.5);
\vertex[particle] (d) at (-2,1.6){$\chi$};
\vertex[particle] (e) at (2,1.6){$\chi$};
\vertex (f) at (0,-2);
\vertex[particle] (g) at (-2,-3.6){$q$};
\vertex[particle] (h) at (2,-3.6){$q$};
\propag[sca] (a) to (d);\propag[sca] (a) to (e);
\propag[sca] (a) to [edge label=$h_j$] (b);\propag[sca] (c) to [edge label=$h_i$] (f);
\propag[fer] (g) to (f); \propag[fer] (f) to (h);
\propag[plain] (b) to [half left, looseness=1.5, edge label=$A$] (c);
\propag[plain] (b) to [half right, looseness=1.5, edge label'=$A$] (c);
\end{feynhand}
\end{tikzpicture}
\subcaption{~~~~(a)}
\end{minipage}
\begin{minipage}[b]{0.2\linewidth}
\centering
\begin{tikzpicture}[scale=0.8]
\begin{feynhand}
\vertex (a) at (0,0);
\vertex (b) at (-1,0.8);
\vertex [ringblob] (beta) at (-1,2) {\scriptsize{$\chi{,}h_k{,}A$}};
\vertex (c) at (-1,2.5);
\vertex[particle] (d) at (-2,1.6){$\chi$};
\vertex[particle] (e) at (2,1.6){$\chi$};
\vertex (f) at (0,-2);
\vertex[particle] (g) at (-2,-3.6){$q$};
\vertex[particle] (h) at (2,-3.6){$q$};
\propag[sca] (a) to [edge label=$\chi$] (b);\propag[sca] (a) to (d);\propag[sca] (a) to (e);
\propag[sca] (a) to [edge label=$h_i$] (f);
\propag[fer] (g) to (f); \propag[fer] (f) to (h);
\propag[sca] (b) to [edge label'=$h_j$] (beta);
\end{feynhand}
\end{tikzpicture}
\subcaption{~~~~(b)}
\end{minipage}
\begin{minipage}[b]{0.2\linewidth}
\centering
\begin{tikzpicture}[scale=0.8]
\begin{feynhand}
\vertex (a) at (0,0);
\vertex (b) at (1,0.8);
\vertex [ringblob] (beta) at (1,2) {\scriptsize{$\chi{,}h_k{,}A$}};
\vertex (c) at (1,2.5);
\vertex[particle] (d) at (-2,1.6){$\chi$};
\vertex[particle] (e) at (2,1.6){$\chi$};
\vertex (f) at (0,-2);
\vertex[particle] (g) at (-2,-3.6){$q$};
\vertex[particle] (h) at (2,-3.6){$q$};
\propag[sca] (a) to [edge label'=$\chi$] (b);\propag[sca] (a) to (d);\propag[sca] (a) to (e);
\propag[sca] (a) to [edge label=$h_i$] (f);
\propag[fer] (g) to (f); \propag[fer] (f) to (h);
\propag[sca] (b) to [edge label=$h_j$] (beta);
\end{feynhand}
\end{tikzpicture}
\subcaption{~~~~(c)}
\end{minipage}
\begin{minipage}[b]{0.2\linewidth}
\centering
\begin{tikzpicture}[scale=0.8]
\begin{feynhand}
\vertex (a) at (0,0);
\vertex (b) at (-1,0.8);
\vertex (beta) at (-0.8,1.6);
\vertex[particle] (d) at (-2,1.6){$\chi$};
\vertex[particle] (e) at (2,1.6){$\chi$};
\vertex (f) at (0,-2);
\vertex[particle] (g) at (-2,-3.6){$q$};
\vertex[particle] (h) at (2,-3.6){$q$};
\propag[sca] (a) to [edge label=$\chi$] (b);\propag[sca] (a) to (d);\propag[sca] (a) to (e);
\propag[sca] (a) to [edge label=$h_i$] (f);
\propag[fer] (g) to (f); \propag[fer] (f) to (h);
\propag[sca] (b) to [half left, looseness=1.5] (beta);
\propag[sca] (b) to [half right, looseness=1.5, edge label'=$\chi{,}h_j$] (beta);
\end{feynhand}
\end{tikzpicture}
\subcaption{~~~~(d)}
\end{minipage}
\begin{minipage}[b]{0.2\linewidth}
\centering
\begin{tikzpicture}[scale=0.8]
\begin{feynhand}
\vertex (a) at (0,0);
\vertex (b) at (1,0.8);
\vertex (beta) at (0.8,1.6);
\vertex[particle] (d) at (-2,1.6){$\chi$};
\vertex[particle] (e) at (2,1.6){$\chi$};
\vertex (f) at (0,-2);
\vertex[particle] (g) at (-2,-3.6){$q$};
\vertex[particle] (h) at (2,-3.6){$q$};
\propag[sca] (a) to [edge label'=$\chi$] (b);\propag[sca] (a) to (d);\propag[sca] (a) to (e);
\propag[sca] (a) to [edge label=$h_i$] (f);
\propag[fer] (g) to (f); \propag[fer] (f) to (h);
\propag[sca] (b) to [half left, looseness=1.5, edge label=$\chi{,}h_j$] (beta);
\propag[sca] (b) to [half right, looseness=1.5] (beta);
\end{feynhand}
\end{tikzpicture}
\subcaption{~~~~(e)}
\end{minipage}
\begin{minipage}[b]{0.2\linewidth}
\centering
\begin{tikzpicture}[scale=0.8]
\begin{feynhand}
\vertex (a) at (0,0);
\vertex[dot] (b) at (-1.6,1.3);
\vertex[dot] (beta) at (-0.5,0.4);
\vertex[particle] (d) at (-2,1.6){$\chi$};
\vertex[particle] (e) at (2,1.6){$\chi$};
\vertex (f) at (0,-2);
\vertex[particle] (g) at (-2,-3.6){$q$};
\vertex[particle] (h) at (2,-3.6){$q$};
\propag[sca] (a) to (d);\propag[sca] (a) to (e);\propag[sca] (a) to [edge label=$\chi$] (beta);
\propag[sca] (a) to [edge label=$h_i$] (f);
\propag[fer] (g) to (f); \propag[fer] (f) to (h);
\propag[sca] (b) to [half right, looseness=1.5, edge label'=$h_j$] (beta);
\propag[sca] (b) to [edge label=$\chi$] (beta);
\end{feynhand}
\end{tikzpicture}
\subcaption{~~~~(f)}
\end{minipage}
\begin{minipage}[b]{0.2\linewidth}
\centering
\begin{tikzpicture}[scale=0.8]
\begin{feynhand}
\vertex (a) at (0,0);
\vertex[dot] (b) at (1.6,1.3);
\vertex[dot] (beta) at (0.5,0.4);
\vertex[particle] (d) at (-2,1.6){$\chi$};
\vertex[particle] (e) at (2,1.6){$\chi$};
\vertex (f) at (0,-2);
\vertex[particle] (g) at (-2,-3.6){$q$};
\vertex[particle] (h) at (2,-3.6){$q$};
\propag[sca] (a) to (d);\propag[sca] (a) to (e);\propag[sca] (a) to [edge label'=$\chi$] (beta);
\propag[sca] (a) to [edge label=$h_i$] (f);
\propag[fer] (g) to (f); \propag[fer] (f) to (h);
\propag[sca] (b) to [half left, looseness=1.5, edge label=$h_j$] (beta);
\propag[sca] (b) to [edge label'=$\chi$] (beta);
\end{feynhand}
\end{tikzpicture} 
\subcaption{~~~~(g)}
\end{minipage}
\begin{minipage}[b]{0.2\linewidth}
\centering
\begin{tikzpicture}[scale=0.8]
\begin{feynhand}
\vertex (a) at (0,0);
\vertex (b) at (0,-0.5);
\vertex (c) at (0,-1.5);
\vertex[particle] (d) at (-2,1.6){$\chi$};
\vertex[particle] (e) at (2,1.6){$\chi$};
\vertex (f) at (0,-2);
\vertex[particle] (g) at (-2,-3.6){$q$};
\vertex[particle] (h) at (2,-3.6){$q$};
\propag[sca] (a) to (d);\propag[sca] (a) to (e);
\propag[sca] (a) to [edge label=$h_j$] (b);\propag[sca] (c) to [edge label=$h_i$] (f);
\propag[fer] (g) to (f); \propag[fer] (f) to (h);
\propag[sca] (b) to [half left, looseness=1.5, edge label=$\chi$] (c);
\propag[sca] (b) to [half right, looseness=1.5, edge label'=$\chi$] (c);
\end{feynhand}
\end{tikzpicture}
\subcaption{~~~~(h)}
\end{minipage}
\begin{minipage}[b]{0.235\linewidth}
\centering
\begin{tikzpicture}[scale=0.8]
\begin{feynhand}
\vertex (a) at (0,0);
\vertex (b) at (-1,0.8);
\vertex (c) at (1,0.8);
\vertex[particle] (d) at (-2,1.6){$\chi$};
\vertex[particle] (e) at (2,1.6){$\chi$};
\vertex (f) at (0,-2);
\vertex[particle] (g) at (-2,-3.6){$q$};
\vertex[particle] (h) at (2,-3.6){$q$};
\propag[sca] (a) to [edge label=$\chi$] (b);\propag[sca] (a) to [edge label'=$\chi$] (c);\propag[sca] (b) to (d);\propag[sca] (c) to (e);
\propag[sca] (b) to [edge label=$h_j$] (c);\propag[sca] (a) to [edge label=$h_i$] (f);
\propag[fer] (g) to (f); \propag[fer] (f) to (h);
\end{feynhand}
\end{tikzpicture}
\subcaption{~~~(i)}
\end{minipage}
\begin{minipage}[b]{0.25\linewidth}
\centering
\begin{tikzpicture}[scale=0.8]
\begin{feynhand}
\vertex (a) at (0,0);
\vertex[particle] (b) at (-1,1.5){$\chi$};
\vertex (c) at (0,-2);
\vertex (d) at (2,0);
\vertex[particle] (e) at (3,1.5){$\chi$};
\vertex (f) at (2,-2);
\vertex[particle] (g) at (-1,-3.5){$q$};
\vertex[particle] (h) at (3,-3.5){$q$};
\propag[sca] (a) to (b);\propag[sca] (a) to [edge label'=$h_i$] (c);\propag[sca] (a) to [edge label=$\chi$] (d); \propag[sca] (d) to (e);\propag[sca] (d) to [edge label=$h_j$] (f);\propag[fer] (c) to [edge label'=$q$] (f);\propag[fer] (g) to (c);\propag[fer] (f) to (h);
\end{feynhand}
\end{tikzpicture}
\subcaption{~~~(j)}
\end{minipage}
\begin{minipage}[b]{0.25\linewidth}
\centering
\begin{tikzpicture}[scale=0.8]
\begin{feynhand}
\vertex (a) at (0,0);
\vertex[particle] (d) at (-2,1.6){$\chi$};
\vertex[particle] (e) at (2,1.6){$\chi$};
\vertex (f) at (0,-1);
\vertex (g) at (-1,-2.5);
\vertex (h) at (1,-2.5);
\vertex (i) at (-2,-3.2);
\vertex (j) at (2,-3.2);
\propag[sca] (a) to (d);\propag[sca] (a) to (e);
\propag[sca] (a) to [edge label=$h_i$] (f);
\propag[fer] (g) to [edge label=$q$](f); \propag[fer] (f) to [edge label=$q$](h);
\propag[fer] (g) to [edge label'=$q$](h);
\propag[glu] (g) to [edge label=$g$](i);\propag[glu] (h) to [edge label'=$g$](j);
\end{feynhand}
\end{tikzpicture}
\subcaption{~~~(k)}
\end{minipage}
\begin{minipage}[b]{0.25\linewidth}
\centering
\begin{tikzpicture}[scale=0.8]
\begin{feynhand}
\vertex (a) at (0,0);
\vertex[particle] (b) at (-1,1.5){$\chi$};
\vertex (c) at (0,-1);
\vertex (d) at (2,0);
\vertex[particle] (e) at (3,1.5){$\chi$};
\vertex (f) at (2,-1);
\vertex (g) at (-1,-3.5);
\vertex (h) at (3,-3.5);
\vertex (i) at (0,-2);
\vertex (j) at (2,-2);
\propag[sca] (a) to (b);\propag[sca] (a) to [edge label'=$h_i$] (c);\propag[sca] (a) to [edge label=$\chi$] (d); \propag[sca] (d) to (e);\propag[sca] (d) to [edge label=$h_j$] (f);\propag[fer] (c) to [edge label=$q$] (f);
\propag[fer] (i) to [edge label=$q$] (c);\propag[fer] (f) to [edge label=$q$] (j);
\propag[glu] (i) to [edge label=$g$](g);\propag[glu] (j) to [edge label'=$g$](h);\propag[fer] (j) to [edge label=$q$] (i);
\end{feynhand}
\end{tikzpicture}
\subcaption{~~~(l)}
\end{minipage}
\caption{Group-1: 1-loop diagrams including the tree-level structure as a subdiagram.}
\label{type1}
\end{figure}
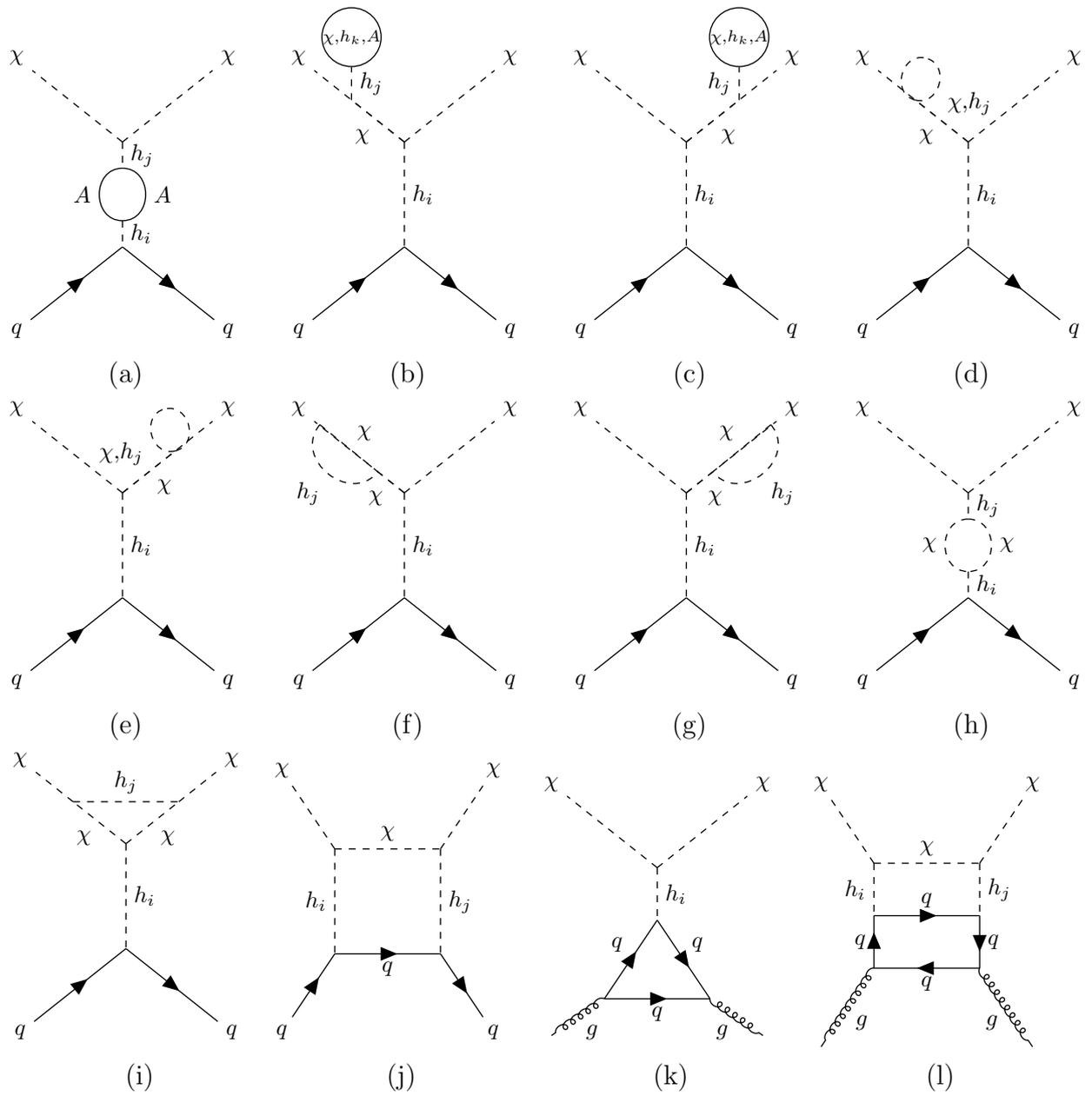
The diagrams in Group-1 (Fig.~\ref{type1}) have the tree-level diagram ($\chi\chi \to (h_i) \to q q$) as a subdiagram. 
Therefore, each diagram in this group vanishes for the same reason as the tree-level. Concerning Fig.~\ref{type1}~(a), particle $A$ represents all the SM particles except for the Higgs, all of which couple only with the SM Higgs $h$ and thus cancel by the same mechanism of the DM-quark scattering in the tree-level.  

\subsection{Group-2}
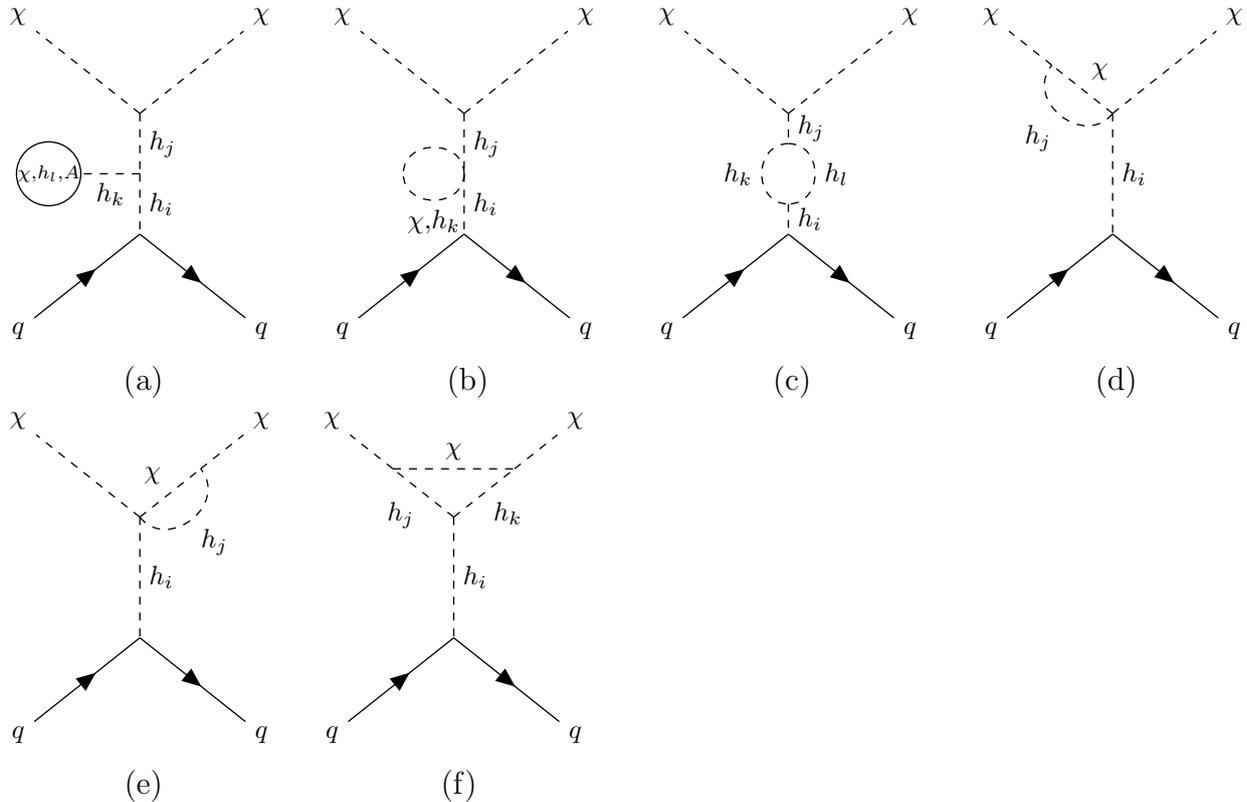
\begin{figure}[h]
\begin{minipage}[b]{0.2\linewidth}
\centering
\begin{tikzpicture}[scale=0.8]
\begin{feynhand}
\vertex (a) at (0,0);
\vertex (b) at (0,-1);
\vertex [ringblob] (c) at (-1.5,-1) {\scriptsize{$\chi{,}h_l{,}A$}};\vertex (gamma) at (-2,-1);
\vertex[particle] (d) at (-2,1.6){$\chi$};
\vertex[particle] (e) at (2,1.6){$\chi$};
\vertex (f) at (0,-2);
\vertex[particle] (g) at (-2,-3.6){$q$};
\vertex[particle] (h) at (2,-3.6){$q$};
\propag[sca] (a) to (d);\propag[sca] (a) to (e);
\propag[sca] (a) to [edge label=$h_j$] (b);\propag[sca] (b) to [edge label=$h_i$] (f);
\propag[fer] (g) to (f); \propag[fer] (f) to (h);
\propag[sca] (b) to [edge label=$h_k$] (c);
\end{feynhand}
\end{tikzpicture}
\subcaption{~~~~(a)}
\end{minipage}
\begin{minipage}[b]{0.2\linewidth}
\centering
\begin{tikzpicture}[scale=0.8]
\begin{feynhand}
\vertex (a) at (0,0);
\vertex (b) at (0,-1);
\vertex (c) at (-1,-1);
\vertex[particle] (d) at (-2,1.6){$\chi$};
\vertex[particle] (e) at (2,1.6){$\chi$};
\vertex (f) at (0,-2);
\vertex[particle] (g) at (-2,-3.6){$q$};
\vertex[particle] (h) at (2,-3.6){$q$};
\propag[sca] (a) to (d);\propag[sca] (a) to (e);
\propag[sca] (a) to [edge label=$h_j$] (b);\propag[sca] (b) to [edge label=$h_i$] (f);
\propag[fer] (g) to (f); \propag[fer] (f) to (h);
\propag[sca] (b) to [half left, looseness=1.5, edge label=$\chi{,}h_k$] (c);
\propag[sca] (b) to [half right, looseness=1.5] (c);
\end{feynhand}
\end{tikzpicture}
\subcaption{~~~~(b)}
\end{minipage}
\begin{minipage}[b]{0.2\linewidth}
\centering
\begin{tikzpicture}[scale=0.8]
\begin{feynhand}
\vertex (a) at (0,0);
\vertex (b) at (0,-0.5);
\vertex (c) at (0,-1.5);
\vertex[particle] (d) at (-2,1.6){$\chi$};
\vertex[particle] (e) at (2,1.6){$\chi$};
\vertex (f) at (0,-2);
\vertex[particle] (g) at (-2,-3.6){$q$};
\vertex[particle] (h) at (2,-3.6){$q$};
\propag[sca] (a) to (d);\propag[sca] (a) to (e);
\propag[sca] (a) to [edge label=$h_j$] (b);\propag[sca] (c) to [edge label=$h_i$] (f);
\propag[fer] (g) to (f); \propag[fer] (f) to (h);
\propag[sca] (b) to [half left, looseness=1.5, edge label=$h_l$] (c);
\propag[sca] (b) to [half right, looseness=1.5, edge label'=$h_k$] (c);
\end{feynhand}
\end{tikzpicture}
\subcaption{~~~~(c)}
\end{minipage}
\begin{minipage}[b]{0.2\linewidth}
\centering
\begin{tikzpicture}[scale=0.8]
\begin{feynhand}
\vertex (a) at (0,0);
\vertex (b) at (-1,0.8);
\vertex[particle] (d) at (-2,1.6){$\chi$};
\vertex[particle] (e) at (2,1.6){$\chi$};
\vertex (f) at (0,-2);
\vertex[particle] (g) at (-2,-3.6){$q$};
\vertex[particle] (h) at (2,-3.6){$q$};
\propag[sca] (a) to [edge label'=$\chi$] (b);\propag[sca] (a) to (e);\propag[sca] (b) to (d);
\propag[sca] (a) to [edge label=$h_i$] (f);
\propag[fer] (g) to (f); \propag[fer] (f) to (h);
\propag[sca] (a) to [half left, looseness=1.5, edge label=$h_j$] (b);
\end{feynhand}
\end{tikzpicture}
\subcaption{~~~~(d)}
\end{minipage}
\begin{minipage}[b]{0.2\linewidth}
\centering
\begin{tikzpicture}[scale=0.8]
\begin{feynhand}
\vertex (a) at (0,0);
\vertex (b) at (1,0.8);
\vertex[particle] (d) at (-2,1.6){$\chi$};
\vertex[particle] (e) at (2,1.6){$\chi$};
\vertex (f) at (0,-2);
\vertex[particle] (g) at (-2,-3.6){$q$};
\vertex[particle] (h) at (2,-3.6){$q$};
\propag[sca] (a) to [edge label=$\chi$] (b);\propag[sca] (a) to (d);\propag[sca] (b) to (e);
\propag[sca] (a) to [edge label=$h_i$] (f);
\propag[fer] (g) to (f); \propag[fer] (f) to (h);
\propag[sca] (a) to [half right, looseness=1.5, edge label'=$h_j$] (b);
\end{feynhand}
\end{tikzpicture}
\subcaption{~~~~(e)}
\end{minipage}
\begin{minipage}[b]{0.33\linewidth}
\centering
\begin{tikzpicture}[scale=0.8]
\begin{feynhand}
\vertex (a) at (0,0);
\vertex (b) at (-1,0.8);
\vertex (c) at (1,0.8);
\vertex[particle] (d) at (-2,1.6){$\chi$};
\vertex[particle] (e) at (2,1.6){$\chi$};
\vertex (f) at (0,-2);
\vertex[particle] (g) at (-2,-3.6){$q$};
\vertex[particle] (h) at (2,-3.6){$q$};
\propag[sca] (a) to [edge label=$h_j$] (b);\propag[sca] (a) to [edge label'=$h_k$] (c);\propag[sca] (b) to (d);\propag[sca] (c) to (e);
\propag[sca] (b) to [edge label=$\chi$] (c);\propag[sca] (a) to [edge label=$h_i$] (f);
\propag[fer] (g) to (f); \propag[fer] (f) to (h);
\end{feynhand}
\end{tikzpicture}
\subcaption{~(f)}
\end{minipage}
\caption{Group-2: 1-loop diagrams including one Yukawa coupling $C_{f f h_i}$ and scalar trilinear coupling $C_{ \chi \chi h_j}$.}
\label{type2}
\end{figure}
The scattering amplitudes described by the diagrams in Fig.~\ref{type2}~($\alpha$), 
where $\alpha=\mathrm{a}\sim \mathrm{f}$, can be expressed as follows\footnote{
The symmetry factor according to the identical particles at the vertex is suppressed.}: 
\begin{align}
  i\mathcal{M}^\alpha 
=
i
\qty(
\sum_{A=1,2}
C_{\chi\chi h_i} C_{qq h_A} F^{\alpha}_{Ai} 
)
f(m^2) \bar{u}(p_3) u(p_1),  
\label{gnrlam1} 
\end{align}
where the scalar $h_A$ is defined as one that couples to the external quark lines, while the scalar with the index $i$ is defined as one that interacts with the $\chi$ pair. 
Besides on $h_A$, second or third scalars appear in the internal line in some diagrams. 
The factor $F^\alpha_{Ai}$ includes all the couplings in the diagrams, and $f(m^2)$ contains propagators. Only exception is Fig.~\ref{type2}~(a), where $F^{\alpha}_{Ai}$ contains only $C_{h_i h_j h_k}$ and tadpole couplings are included in $f(m^2)$. We will show the concrete expression of $F^{\alpha}_{Ai}$ and $f(m^2)$ 
in Fig.~\ref{type2}~(a) soon, and ones for the others in Appendix~\ref{app2}. 
Taking account of Eqs.~\eqref{h1h2yukawa}, \eqref{ccssdd} and \eqref{dlt2}, 
the condition to suppress the scattering amplitude $\qty(\mathcal{M}^\alpha=0)$ is given as follows: 
\begin{align}
\sum_{A=1,2}
C_{\chi\chi h_i} C_{qq h_A} F^{\alpha}_{Ai} 
&=
C_{\chi \chi s} 
\qty{
\qty({F^{\alpha}_{11}}-{F^{\alpha}_{22}})\cos{\alpha}\sin{\alpha} + {F^{\alpha}_{12}}\cos^2\alpha-{F^{\alpha}_{21}}\sin^2\alpha}
\nonumber \\
&=
0. 
\label{type2cond3}
\end{align}
Since $C_{\chi\chi h}$ is given by $\delta_2$ in the pNG-DM model (see, \eqref{c_cch}), 
we have dropped $C_{\chi\chi h}$ because $\delta_2=0$ for $m_{h_1}=m_{h_2}$ \eqref{dlt2}. 
In general, the parameter $F^{\alpha}_{Ai}$ is given by the scalar trilinear and/or quartic couplings, 
which are given by $\delta_2,~\lambda$ and $d_2$ in \eqref{mnmlmdl}.  
Thus, in the degenerate mass limit of two scalars $(\delta_2=0)$, $F^{\alpha}_{Ai}$ can be expanded as 
\begin{align}
F^{\alpha}_{1i}= a_{1i}^\alpha \lambda + b_{1i}^\alpha d_2, 
~~~
F^{\alpha}_{2i}= a_{2i}^\alpha \lambda + b_{2i}^\alpha d_2. 
\label{vertexinF}
\end{align}
Therefore, the condition \eqref{type2cond3} is replaced by the following two conditions, which should be satisfied simultaneously 
\begin{align}
\qty(a_{11}^\alpha -a_{22}^\alpha ) \cos\alpha \sin\alpha
+ a_{12}^\alpha \cos^2 \alpha
- a_{21}^\alpha \sin^2 \alpha
&=
0, 
\label{cnda}
\\
\qty(b_{11}^\alpha -b_{22}^\alpha ) \cos\alpha \sin\alpha
+ b_{12}^\alpha \cos^2 \alpha
- b_{21}^\alpha \sin^2 \alpha
&=
0.  
\label{cndb}
\end{align}

As an example, we examine the amplitude given by Fig.~\ref{type2}~(a): 
\begin{align}
 i\mathcal{M}^\mathrm{a}
&=
\frac{-1}{m^6} 
\bar{u}(p_3) u(p_1)
\qty(
\sum_{l=\chi,h_1,h_2,A} C_{ll h_k} \frac{i}{16\pi^2}A(m_l) 
)
\nonumber \\
&~~~~\times
\qty(
C_{qq h_1} C_{\chi\chi h_j} C_{h_1 h_j h_k}
+
C_{qq h_2} C_{\chi\chi h_j} C_{h_2 h_j h_k}
), 
\label{ampf4a}
\end{align}
where $j,k=1,2$, and $A(m)$ is the $A$-function by Passarino and Veltman~\cite{Passarino:1978jh}. 
Comparing \eqref{ampf4a} with \eqref{gnrlam1}, the parameter $F_{Ai}^\alpha$ is given by the trilinear couplings as 
\begin{align}
F^\mathrm{a}_{1j}=C_{h_1 h_j h_k},~~~F^\mathrm{a}_{2j}=C_{h_2 h_j h_k}. 
\end{align}
The trilinear and quartic couplings in the pNG-DM model are summarized in Appendix~\ref{app1}. 
Taking account of the symmetry factors, $F_{Ai}^\mathrm{a}$ for $k=1$ are given as 
\begin{align}
F^\mathrm{a}_{11}= 6 C_{111}&= \frac{3}{2} \qty( \lambda v \cos^3 \alpha + d_2 v_s \sin^3 \alpha), 
\label{d2sfa}
\\
F^\mathrm{a}_{12}= 2 C_{112}&= \frac{3}{2} \qty(-\lambda v \sin{\alpha}\cos^2{\alpha}+ d_2 v_s \sin^2{\alpha}\cos{\alpha}),  \\
F^\mathrm{a}_{21}= 2 C_{112}&=\frac{3}{2} \qty(-\lambda v \sin{\alpha}\cos^2{\alpha}+ d_2 v_s \sin^2{\alpha}\cos{\alpha}), \\
F^\mathrm{a}_{22}= 2 C_{122}&=\frac{3}{2} \qty( \lambda v \sin^2{\alpha}\cos{\alpha}+ d_2 v_s \sin{\alpha}\cos^2{\alpha}).  
\label{d2sfe}
\end{align}
It is easy to confirm that 
Eqs.~\eqref{d2sfa}-\eqref{d2sfe}, and those for $k=2$ satisfy the conditions \eqref{cnda} and \eqref{cndb}, 
so the amplitude given by Fig.~\ref{type2}~(a) vanishes in the degenerate scalar limit. 
It can be shown that the other diagrams in Fig.~\ref{type2} also vanish~(See Appendix \ref{app2} for detail).

\subsection{Group-3}\label{sec;g3}
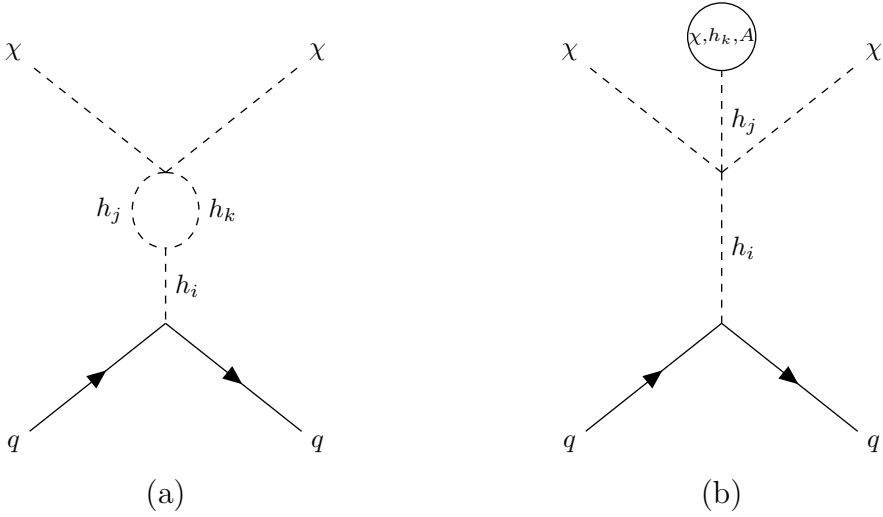
\begin{figure}[htpb]
\begin{minipage}[b]{0.45\linewidth}
\centering
\begin{tikzpicture}
\begin{feynhand}
\vertex (a) at (0,0);
\vertex (b) at (0,-1);
\vertex[particle] (d) at (-2,1.6){$\chi$};
\vertex[particle] (e) at (2,1.6){$\chi$};
\vertex (f) at (0,-2);
\vertex[particle] (g) at (-2,-3.6){$q$};
\vertex[particle] (h) at (2,-3.6){$q$};
\propag[sca] (a) to (d);\propag[sca] (a) to (e);
\propag[sca] (b) to [edge label=$h_i$] (f);
\propag[fer] (g) to (f); \propag[fer] (f) to (h);
\propag[sca] (a) to [half left, looseness=1.5, edge label=$h_k$] (b);
\propag[sca] (a) to [half right, looseness=1.5, edge label'=$h_j$] (b);
\end{feynhand}
\end{tikzpicture}
\subcaption{(a)}
\end{minipage}
\begin{minipage}[b]{0.45\linewidth}
\centering
\begin{tikzpicture}
\begin{feynhand}
\vertex (a) at (0,0);
\vertex [ringblob] (b) at (0,1.8) {\scriptsize{$\chi{,}h_k{,}A$}};
\vertex (c) at (0,2.5);
\vertex[particle] (d) at (-2,1.6){$\chi$};
\vertex[particle] (e) at (2,1.6){$\chi$};
\vertex (f) at (0,-2);
\vertex[particle] (g) at (-2,-3.6){$q$};
\vertex[particle] (h) at (2,-3.6){$q$};
\propag[sca] (a) to (d);\propag[sca] (a) to (e);
\propag[sca] (a) to [edge label=$h_i$] (f);
\propag[fer] (g) to (f); \propag[fer] (f) to (h);
\propag[sca] (a) to [edge label'=$h_j$]  (b);
\end{feynhand}
\end{tikzpicture}
\subcaption{(b)}
\end{minipage}
\caption{Group-3: 1-loop diagrams including one Yukawa coupling $C_{ f f h_i}$ and scalar quartic coupling $C_{\chi \chi h_j h_k }$.}
\label{type3}
\end{figure}
The diagrams in Group-3  have the Yukawa couplings $C_{qq h_i}$ and the scalar quartic coupling $C_{h_i h_j \chi \chi}$ as shown 
in Figs.~\ref{type3}~(a) and \ref{type3}~(b). 
The amplitude of Fig.~\ref{type3}~(a) is given as
\begin{align}
 i\mathcal{M}^\mathrm{a}
&=
\frac{1}{m^2}\frac{i}{16\pi^2}B_0(m, m) \bar{u}(p_3) u(p_1) 
\qty(
C_{qq h_1} C_{h_1 h_j h_k} + C_{qq h_2} C_{h_2 h_j h_k} 
)C_{\chi\chi h_j h_k}, 
\end{align}
where $B_0(m, m)$ is the $B$-function defined in \cite{Passarino:1978jh}. 
The amplitude vanishes when the following relation holds: 
\begin{align}
\qty(
C_{h_1 h_j h_k} \cos\alpha - C_{h_2 h_j h_k} \sin\alpha
)C_{\chi\chi h_j h_k} 
= 0, 
\label{rule_gr3_1}
\end{align}
where \eqref{h1h2yukawa} is used. 
Summing up \eqref{rule_gr3_1} for $j$ and $k$, and substituting trilinear and quartic couplings given in 
Appendix~\ref{app1}, we can confirm that \eqref{rule_gr3_1} is satisfied. 
Next, We consider Fig.~\ref{type3}~(b). The scattering amplitude is given by 
\begin{align}
 i\mathcal{M}^\mathrm{b}
&=
\frac{-1}{m^4} \bar{u}(p_3) u(p_1) 
\qty(
\sum_{l=\chi,h_1,h_2,A} C_{ll A} \frac{i}{16\pi^2} A(m_l)
)
\qty(
C_{qq h_1} C_{\chi\chi h_j h_1 }
+C_{qq h_2} C_{\chi\chi h_j h_2}
). 
\end{align}
The sum rule for $\mathcal{M}^\mathrm{b}=0$ is 
\begin{align}
C_{\chi\chi h_j h_1 }\cos\alpha
- C_{\chi\chi h_j h_2}\sin\alpha
 =0, 
\label{rule_gr3_2}
\end{align}
and it is straightforward to confirm this for both $j=1$ and $2$ in the pNG-DM model. 

\subsection{Group-4}\label{sec;g4}
\begin{figure}[htbp]
\begin{minipage}[h]{1\linewidth}
\centering
\begin{tikzpicture}
\begin{feynhand}
\vertex (a) at (0,0);
\vertex[particle] (d) at (-2,1.6){$\chi$};
\vertex[particle] (e) at (2,1.6){$\chi$};
\vertex (f) at (0,-2);
\vertex[particle] (g) at (-2,-3.6){$q$};
\vertex[particle] (h) at (2,-3.6){$q$};
\vertex (i) at (-1,-2.8);
\vertex (j) at (1,-2.8);
\propag[sca] (a) to (d);\propag[sca] (a) to (e);
\propag[sca] (a) to [edge label=$h_i$] (f);
\propag[sca] (f) to [edge label'=$h_j$] (i); \propag[sca] (f) to [edge label=$h_k$] (j);
\propag[fer] (g) to (i); \propag[fer] (j) to (h);
\propag[fer] (i) to [edge label=$q$] (j);
\end{feynhand}
\end{tikzpicture}
\end{minipage}
\caption{Group-4: 1-loop diagram including two Yukawa coupling $C_{ f f h_i}$ and scalar trilinear coupling $C_{\chi \chi h_j }$.}
\label{type4}
\end{figure}
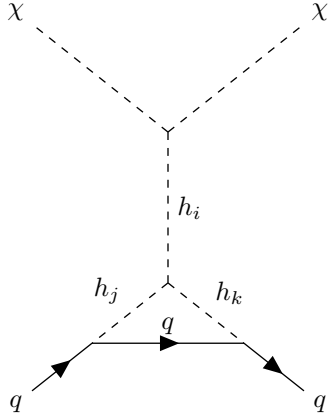
The Feynman diagram of Group-4 given in Fig.~(\ref{type4}) has two Yukawa couplings. 
The amplitude is given by
\begin{align}
 i\mathcal{M}
&=
\frac{-1}{m^2} \bar{u}(p_3) u(p_1) 
\qty(
\frac{i}{16\pi^2} C^\mu (m_q, m, m)
)
\nonumber \\
&~~~~~
\times 
\qty(
C_{\chi \chi h_1} C_{h_1 h_j h_k}
+
C_{\chi \chi h_2} C_{h_2 h_j h_k}
)
C_{qq h_j}C_{qq h_k}, 
\end{align}
where the rank-1 $C$-function $C^\mu$ is defined in \cite{Passarino:1978jh}. 
The sum rule for $\mathcal{M}=0$ is 
\begin{align}
\qty(
C_{\chi \chi h_1} C_{h_1 h_j h_k}
+
C_{\chi \chi h_2} C_{h_2 h_j h_k}
)
C_{qq h_j}C_{qq h_k} 
&= 0.  
\label{rule_gr4}
\end{align}

\subsection{Group-5}\label{sec;g5}
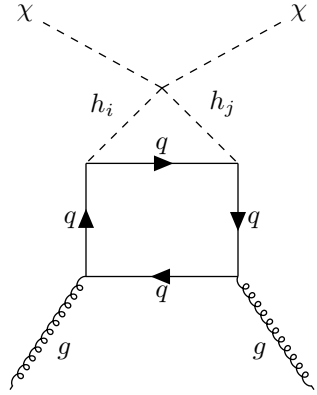
\begin{figure}[htbp]
\begin{minipage}[h]{1\linewidth}
\centering
\begin{tikzpicture}
\begin{feynhand}
\vertex (a) at (1,0.5);
\vertex[particle] (b) at (-0.8,1.5){$\chi$};
\vertex (c) at (0,-0.5);
\vertex (d) at (2,0);
\vertex[particle] (e) at (2.8,1.5){$\chi$};
\vertex (f) at (2,-0.5);
\vertex (g) at (-1,-3.5);
\vertex (h) at (3,-3.5);
\vertex (i) at (0,-2);
\vertex (j) at (2,-2);
\propag[sca] (a) to (b); \propag[sca] (a) to (e);\propag[sca] (a) to [edge label'=$h_i$](c); \propag[sca] (a) to [edge label=$h_j$](f);
\propag[fer] (c) to [edge label=$q$] (f);
\propag[fer] (i) to [edge label=$q$] (c);\propag[fer] (f) to [edge label=$q$] (j);
\propag[glu] (i) to [edge label=$g$](g);\propag[glu] (j) to [edge label'=$g$](h);\propag[fer] (j) to [edge label=$q$] (i);
\end{feynhand}
\end{tikzpicture}
\end{minipage}
\caption{Group-5: 1-loop diagram including two Yukawa coupling $C_{ f f h_i}$ and scalar quartic coupling $C_{\chi \chi h_j h_k }$.}
\label{type5}
\end{figure}
The Feynman diagram of Group-5 is given in Fig.~\ref{type5}, which has the scalar quartic coupling and two Yukawa couplings. 
Up to the quark-gluon scattering part, 
the amplitudes of Fig.~\ref{type5} is given by
\begin{align}
 i\mathcal{M}
&=
\qty(\frac{i}{16\pi^2} C^\mu (m_q, m, m)) \bar{u}(p_3) u(p_1)
C_{\chi\chi h_i h_j} C_{qq h_i} C_{qq h_j}. 
\end{align}
The sum rule for $\mathcal{M}=0$ is 
\begin{align}
C_{\chi\chi h_i h_j} C_{qq h_i} C_{qq h_j} =0. 
\label{rule_gr5}
\end{align}

Although we did not mention in Sec.~\ref{sec;g3}, \ref{sec;g4} and \ref{sec;g5}, 
we confirmed that the sum rules \eqref{rule_gr3_1}, \eqref{rule_gr3_2}, \eqref{rule_gr4} and \eqref{rule_gr5} are satisfied in the pNG-DM models using the couplings summarized in Appendix.\ref{app1}. 

\section{Summary}\label{sec;sum}

We have investigated the origin of the cancellation mechanism of the DM-quark scattering processes in the degenerate limit of mediator scalars ($h_1$ and $h_2$) in their masses, using the general scalar potential of the CxSM. 
Our analysis at the tree-level revealed that the relationship between scalar trilinear and bilinear couplings, which describe the mixing of the SM Higgs and the singlet scalar, is crucial for the cancellation. 
Therefore, the cancellation mechanism restricts some operators in the general scalar potential in the CxSM. 
We found a sum rule for the couplings in the scalar potential to guarantee the cancellation of the amplitudes. It was shown that the pNG-DM model is an example that satisfies the tree-level sum rule. 
We have also extended our study to the 1-loop level. 
Many 1-loop diagrams were classified into five groups, and the sum rules for the couplings were derived in each group. Since it was already shown that the general CxSM has no capability of the cancellation mechanism even at the tree-level, we performed that the sum rules at the 1-loop level were satisfied in the pNG-DM model in the limit of the degenerate scalars in their masses. 
It would be interesting to see if the suppression mechanism found in this study can be applied to other scalar DM models with the Higgs portal, e.g., an extension of the two-Higgs doublet model with a singlet pseudoscalar~\cite{vonBuddenbrock:2016rmr}. This is left for future work.

\section*{Acknowledgements}
The work of G.C.C. is supported in part by JSPS KAKENHI Grant No. 22K03616. The work was also partly supported by MEXT Promotion of Distinctive Joint Research Center Program JPMXP0619217849.

\newpage
\appendix
\section{Trilinear and quartic couplings in the pNG-DM model}\label{app1}
We collect trilinear and quartic couplings of the scalar potential in the pNG-DM model, which need to confirm the cancellation of the DM-quark scatterings in the 1-loop level discussed in Sec.~\ref{sec;loop}. 
In the mass eigenstates of scalars $h_1,h_2$ and the DM $\chi$, 
the trilinear and quartic terms of the scalar potential are given by
\begin{align}
V = & C_{111} h_1^3+C_{112} h_1^2 h_2+C_{122} h_1 h_2^2+C_{222} h_2^3+C_{\chi \chi 1} h_1 \chi^2+C_{\chi \chi 2} h_2 \chi^2 \nonumber \\
& +C_{1111} h_1^4+C_{1112} h_1^3 h_2+C_{1122} h_1^2 h_2^2+C_{1222} h_1 h_2^3+C_{2222} h_2^4 \nonumber \\
& +C_{\chi \chi 11} h_1^2 \chi^2+C_{\chi \chi 12} h_1 h_2 \chi^2+C_{\chi \chi 22} h_2^2 \chi^2+C_{\chi \chi \chi \chi} \chi^4.  
\end{align}
The explicit forms of the couplings are summarized as follows: 
\begin{align}
C_{111}&=\frac{\lambda}{4}v \cos^3{\alpha}+\frac{\delta_2}{4}v \sin^2{\alpha}\cos{\alpha}+\frac{\delta_2}{4}v_S \sin{\alpha}\cos^2{\alpha}+\frac{d_2}{4}v_s \sin^3{\alpha},\\
C_{112}&=-\frac{3 \lambda}{4}v \sin{\alpha}\cos^2{\alpha}+\frac{\delta_2}{4}v (2\sin{\alpha}\cos^2{\alpha}-\sin^3{\alpha})\nonumber, \\
&+\frac{\delta_2}{4}v_S (\cos^3{\alpha} - 2\sin^2{\alpha}\cos{\alpha})+\frac{3 d_2}{4}v_s \sin^2{\alpha}\cos{\alpha}, \\
C_{122}&=\frac{3 \lambda}{4}v \sin^2{\alpha}\cos{\alpha}+\frac{\delta_2}{4}v (\cos^3{\alpha} - 2\sin^2{\alpha}\cos{\alpha})\nonumber \\
&-\frac{\delta_2}{4}v_S (2\sin{\alpha}\cos^2{\alpha}-\sin^3{\alpha})+\frac{3 d_2}{4}v_s \sin{\alpha}\cos^2{\alpha}, \\
C_{222}&=-\frac{\lambda}{4}v \sin^3{\alpha}-\frac{\delta_2}{4}v \sin{\alpha}\cos^2{\alpha}+\frac{\delta_2}{4}v_S \sin^2{\alpha}\cos{\alpha}+\frac{d_2}{4}v_s \cos^3{\alpha}, \\
C_{\chi \chi 1}&=\frac{\delta_2}{4}v \cos{\alpha}+\frac{d_2}{4}v_S \sin{\alpha}, \\
C_{\chi \chi 2}&=-\frac{\delta_2}{4}v \sin{\alpha}+\frac{d_2}{4}v_S \cos{\alpha}, \\
C_{1111}&=\frac{\lambda}{16}\cos^4{\alpha}+\frac{\delta_2}{8}\sin^2{\alpha}\cos^2{\alpha}+\frac{d_2}{16}\sin^4{\alpha}, \\
C_{1112}&=-\frac{\lambda}{4}\sin{\alpha}\cos{\alpha}^3+\frac{\delta_2}{4}(\sin{\alpha}\cos{\alpha}^3-\sin{\alpha}^3\cos{\alpha})+\frac{d_2}{4}\sin^3{\alpha}\cos{\alpha}, \\
C_{1122}&=\frac{3\lambda}{8}\sin^2{\alpha}\cos^2{\alpha}+\frac{\delta_2}{8}(\sin^4{\alpha}+\cos^4{\alpha}-4\sin^2{\alpha}\cos^2{\alpha})+\frac{3 d_2}{8}\sin^2{\alpha}\cos^2{\alpha}, \\
C_{1222}&=-\frac{\lambda}{4}\sin^3{\alpha}\cos{\alpha}+\frac{\delta_2}{4}(-\sin{\alpha}\cos^3{\alpha}+\sin^3{\alpha}\cos{\alpha})+\frac{d_2}{4}\sin{\alpha}\cos^3{\alpha}, \\
C_{2222}&=\frac{\lambda}{16}\sin^4{\alpha}+\frac{\delta_2}{8}\sin^2{\alpha}\cos^2{\alpha}+\frac{d_2}{16}\cos^4{\alpha},  \\
C_{\chi \chi 11}&=\frac{\delta_2}{8}\cos^2{\alpha}+\frac{d_2}{8}\sin^2{\alpha}, \\
C_{\chi \chi 12}&=-\frac{\delta_2}{4}\sin{\alpha}\cos{\alpha}+\frac{d_2}{4}\sin{\alpha}\cos{\alpha}, \\
C_{\chi \chi 22}&=\frac{\delta_2}{8}\sin^2{\alpha}+\frac{d_2}{8}\cos^2{\alpha}, \\
C_{\chi \chi \chi \chi}&=\frac{d_2}{16}.
\end{align}

\section{Amplitudes of 1-loop diagrams in Group-2}\label{app2}
\subsection{Fig.~\ref{type2}~(b) diagram}
The scattering amplitudes of Fig.~\ref{type2}~(b) can be expressed as
\begin{align}
 i\mathcal{M}^\mathrm{b}
&=
\frac{-1}{m^4} 
\bar{u}(p_3) u(p_1)
\qty(
\sum_{l=\chi,h_1,h_2}  \frac{i}{16\pi^2}A(m_l) 
)
\nonumber \\
&~~~~\times
\qty(
C_{qq h_1} C_{\chi\chi h_j} C_{ll h_1 h_j}
+
C_{qq h_2} C_{\chi\chi h_j} C_{ll h_2 h_k}
).
\end{align}
The following functions can be obtained by comparing with Eq.~\eqref{gnrlam1};  
\begin{align}
F^\mathrm{b}_{1j}=\left\{\begin{array}{l}
C_{\chi\chi h_1 h_j}~~(\mathrm{for}~~\chi\mathrm{\mathchar`-loop}) \\
C_{h_1h_1h_1h_j}~~(\mathrm{for}~~h_1\mathrm{\mathchar`-loop})\\
C_{h_2h_2h_1h_j}~~(\mathrm{for}~~h_2\mathrm{\mathchar`-loop})
\end{array}\right.,~~~F^\mathrm{b}_{2j}=\left\{\begin{array}{l}
C_{\chi\chi h_2 h_j}~~(\mathrm{for}~~\chi\mathrm{\mathchar`-loop}) \\
C_{h_1h_1h_2h_j}~~(\mathrm{for}~~h_1\mathrm{\mathchar`-loop})\\
C_{h_2h_2h_2h_j}~~(\mathrm{for}~~h_2\mathrm{\mathchar`-loop})
\end{array}\right. .
\end{align}
The explicit form of couplings $F^\mathrm{b}_{1j}$ and $F^\mathrm{b}_{2j}$ in each loop are summarized as follows: 
\begin{itemize}
 \item $\chi$-loop
\begin{align}
&F^\mathrm{b}_{11}=2  C_{\chi\chi 1 1},~~~F^\mathrm{b}_{12}=C_{\chi\chi 1 2},\nonumber \\
&F^\mathrm{b}_{21}=C_{\chi\chi 1 2},~~~F^\mathrm{b}_{22}=2  C_{\chi\chi 2 2}, 
\end{align}
\item $h_1$-loop
\begin{align}
&F^\mathrm{b}_{11}=12  C_{1111},~~~F^\mathrm{b}_{12}=3 C_{1112},\nonumber \\
&F^\mathrm{b}_{21}=3 C_{1112},~~~F^\mathrm{b}_{22}=2 C_{1122},
\end{align}
\item $h_2$-loop
\begin{align}
&F^\mathrm{b}_{11}=2 C_{1122},~~~F^\mathrm{b}_{12}=3 C_{1222},\nonumber \\
&F^\mathrm{b}_{21}=3 C_{1222},~~~F^\mathrm{b}_{22}=12 C_{2222}.
\end{align}
\end{itemize}

\subsection{Fig.~\ref{type2}~(c) diagram}
The scattering amplitudes of Fig.~\ref{type2}~(c) is given as 
\begin{align}
 i\mathcal{M}^\mathrm{c}
&=
\frac{1}{m^4}
\bar{u}(p_3) u(p_1)
\qty(
\frac{i}{16\pi^2}B_0(m,m) 
)
\nonumber \\
&~~~~\times
\qty(
C_{qq h_1} C_{\chi\chi h_j} C_{h_1h_kh_l}C_{h_jh_kh_l}
+
C_{qq h_2} C_{\chi\chi h_j} C_{h_2h_kh_l}C_{h_jh_kh_l}
).
\end{align}
The following functions can be obtained by comparing with Eq.~\eqref{gnrlam1}; 
\begin{align}
F^\mathrm{c}_{1j}=C_{h_1h_kh_l}C_{h_jh_kh_l},~~~F^\mathrm{c}_{2j}=C_{h_2h_kh_l}C_{h_jh_kh_l}.
\end{align}
The explicit form of couplings $F^\mathrm{c}_{1j}$ and $F^\mathrm{c}_{2j}$ in each loop are summarized as follows:
\begin{itemize}
\item  $(h_k,h_l)=(h_1,h_1)$
\begin{align}
&F^\mathrm{c}_{11}=18 C_{111}C_{111},~~~F^\mathrm{c}_{21}=6  C_{112}C_{111},\nonumber \\
&F^\mathrm{c}_{12}=6 C_{111}C_{112},~~~F^\mathrm{c}_{22}=2 C_{112}C_{112},
\end{align}
\item $(h_k,h_l)=(h_1,h_2)$, $(h_2,h_1)$
\begin{align}
&F^\mathrm{c}_{11}=2 C_{112}C_{112},~~~F^\mathrm{c}_{21}=2  C_{122}C_{112},\nonumber \\
&F^\mathrm{c}_{12}=2  C_{112}C_{122},~~~F^\mathrm{c}_{22}=2  C_{122}C_{122},
\end{align}
\item $(h_k,h_l)=(h_2,h_2)$
\begin{align}
&F^\mathrm{c}_{11}=2 C_{122}C_{122},~~~F^\mathrm{c}_{21}=6  C_{222}C_{122},\nonumber \\
&F^\mathrm{c}_{12}=6  C_{122}C_{222},~~~F^\mathrm{c}_{22}=18  C_{222}C_{222}.
\end{align}
\end{itemize}

\subsection{Fig.~\ref{type2}~(d) and Fig.~\ref{type2}~(e) diagram}
The scattering amplitudes of Fig.~\ref{type2}~(d) and Fig.~\ref{type2}~(e) can be expressed as
\begin{align}
 i\mathcal{M}^{de}
&=
\frac{-1}{m^2}
\bar{u}(p_3) u(p_1)
\qty(
\frac{i}{16\pi^2}B_0(m,m_{\chi}) 
)
\nonumber \\
&~~~~\times
\qty(
C_{qq h_1} C_{\chi\chi h_j} C_{\chi\chi h_1 h_j}
+
C_{qq h_2} C_{\chi\chi h_j} C_{\chi\chi h_2 h_j}
).
\end{align}
The following functions can be obtained by comparing with Eq.~\eqref{gnrlam1};
\begin{align}
F^\mathrm{de}_{1j}=C_{\chi\chi h_1h_j},~~~F^\mathrm{de}_{2j}=C_{\chi\chi h_2h_j}.
\end{align}
The explicit form of couplings $F^\mathrm{de}_{1j}$ and $F^\mathrm{de}_{2j}$ in each loop are summarized as follows:
\begin{itemize}
\item $j=1$
\begin{align}
&F^\mathrm{de}_{11}=4 C_{\chi\chi11},~~~F^\mathrm{de}_{12}=2 C_{\chi\chi12},
\end{align}
\item $j=2$
\begin{align}
&F^\mathrm{de}_{21}=2 C_{\chi\chi12},~~~F^\mathrm{de}_{22}=4 C_{\chi\chi22}.
\end{align}
\end{itemize}

\subsection{Fig.~\ref{type2}~(f) diagram}
The scattering amplitudes of Fig.~\ref{type2}~(f) can be expressed as
\begin{align}
 i\mathcal{M}^\mathrm{f}
&=
\frac{1}{m^2}
\bar{u}(p_3) u(p_1)
\qty(
\frac{i}{16\pi^2}C_0(m_{\chi},m,m) 
)
\nonumber \\
&~~~~\times
\qty(
C_{qq h_1} C_{\chi\chi h_j} C_{\chi\chi h_k} C_{h_1 h_j h_k}
+ 
C_{qq h_2} C_{\chi\chi h_j} C_{\chi\chi h_k} C_{h_2 h_j h_k}
).
\end{align}
The following functions can be obtained by comparing with Eq.~\eqref{gnrlam1};
\begin{align}
F^\mathrm{f}_{1j}=C_{\chi\chi h_k} C_{h_1 h_j h_k},~~~F^\mathrm{f}_{2j}=C_{\chi\chi h_k} C_{h_1 h_j h_k}.
\end{align}
The explicit form of couplings $F^\mathrm{f}_{1j}$ and $F^\mathrm{f}_{2j}$ in each loop are summarized as follows:
\begin{itemize}
\item $k=1$
\begin{align}
&F^\mathrm{f}_{11}=12 C_{\chi\chi 1}C_{111},~~~F^\mathrm{f}_{12}=4 C_{\chi\chi 1}C_{112},\nonumber \\
&F^\mathrm{f}_{21}=4 C_{\chi\chi 1}C_{112},~~~F^\mathrm{f}_{22}=4 C_{\chi\chi 1}C_{122}.
\end{align}
\item $k=2$
\begin{align}
&F^\mathrm{f}_{11}=4 C_{\chi\chi 2}C_{112},~~~F^\mathrm{f}_{12}=4 C_{\chi\chi 2}C_{122},\nonumber \\
&F^\mathrm{f}_{21}=4 C_{\chi\chi 2}C_{122},~~~F^\mathrm{f}_{22}=12 C_{\chi\chi 2}C_{222}.
\end{align}
\end{itemize}

\bibliography{biblist}

\providecommand{\href}[2]{#2}\begingroup\raggedright\begin{thebibliography}{10}

\bibitem{LZ:2022ufs}
{\scshape LZ} collaboration, J.~Aalbers et~al., \emph{{First Dark Matter Search
  Results from the LUX-ZEPLIN (LZ) Experiment}},
  \href{https://arxiv.org/abs/2207.03764}{{\ttfamily 2207.03764}}.

\bibitem{Ipek:2014gua}
S.~Ipek, D.~McKeen and A.~E. Nelson, \emph{{A Renormalizable Model for the
  Galactic Center Gamma Ray Excess from Dark Matter Annihilation}},
  \href{https://doi.org/10.1103/PhysRevD.90.055021}{\emph{Phys. Rev. D}
  {\bfseries 90} (2014) 055021},
  [\href{https://arxiv.org/abs/1404.3716}{{\ttfamily 1404.3716}}].

\bibitem{Escudero:2016gzx}
M.~Escudero, A.~Berlin, D.~Hooper and M.-X. Lin, \emph{{Toward (Finally!)
  Ruling Out Z and Higgs Mediated Dark Matter Models}},
  \href{https://doi.org/10.1088/1475-7516/2016/12/029}{\emph{JCAP} {\bfseries
  12} (2016) 029}, [\href{https://arxiv.org/abs/1609.09079}{{\ttfamily
  1609.09079}}].

\bibitem{Abe:2018emu}
T.~Abe, M.~Fujiwara and J.~Hisano, \emph{{Loop corrections to dark matter
  direct detection in a pseudoscalar mediator dark matter model}},
  \href{https://doi.org/10.1007/JHEP02(2019)028}{\emph{JHEP} {\bfseries 02}
  (2019) 028}, [\href{https://arxiv.org/abs/1810.01039}{{\ttfamily
  1810.01039}}].

\bibitem{Gross:2017dan}
C.~Gross, O.~Lebedev and T.~Toma, \emph{{Cancellation Mechanism for
  Dark-Matter\textendash{}Nucleon Interaction}},
  \href{https://doi.org/10.1103/PhysRevLett.119.191801}{\emph{Phys. Rev. Lett.}
  {\bfseries 119} (2017) 191801},
  [\href{https://arxiv.org/abs/1708.02253}{{\ttfamily 1708.02253}}].

\bibitem{Chao:2017vrq}
W.~Chao, H.-K. Guo and J.~Shu, \emph{{Gravitational Wave Signals of Electroweak
  Phase Transition Triggered by Dark Matter}},
  \href{https://doi.org/10.1088/1475-7516/2017/09/009}{\emph{JCAP} {\bfseries
  09} (2017) 009}, [\href{https://arxiv.org/abs/1702.02698}{{\ttfamily
  1702.02698}}].

\bibitem{Ghorbani:2018pjh}
K.~Ghorbani and P.~H. Ghorbani, \emph{{Leading Loop Effects in
  Pseudoscalar-Higgs Portal Dark Matter}},
  \href{https://doi.org/10.1007/JHEP05(2019)096}{\emph{JHEP} {\bfseries 05}
  (2019) 096}, [\href{https://arxiv.org/abs/1812.04092}{{\ttfamily
  1812.04092}}].

\bibitem{Jiang:2019soj}
X.-M. Jiang, C.~Cai, Z.-H. Yu, Y.-P. Zeng and H.-H. Zhang,
  \emph{{Pseudo-Nambu-Goldstone dark matter and two-Higgs-doublet models}},
  \href{https://doi.org/10.1103/PhysRevD.100.075011}{\emph{Phys. Rev. D}
  {\bfseries 100} (2019) 075011},
  [\href{https://arxiv.org/abs/1907.09684}{{\ttfamily 1907.09684}}].

\bibitem{Abe:2021nih}
S.~Abe, G.-C. Cho and K.~Mawatari, \emph{{Probing a degenerate-scalar scenario
  in a pseudoscalar dark-matter model}},
  \href{https://doi.org/10.1103/PhysRevD.104.035023}{\emph{Phys. Rev. D}
  {\bfseries 104} (2021) 035023},
  [\href{https://arxiv.org/abs/2101.04887}{{\ttfamily 2101.04887}}].

\bibitem{Cai:2021evx}
C.~Cai, Y.-P. Zeng and H.-H. Zhang, \emph{{Cancellation mechanism of dark
  matter direct detection in Higgs-portal and vector-portal models}},
  \href{https://doi.org/10.1007/JHEP01(2022)117}{\emph{JHEP} {\bfseries 01}
  (2022) 117}, [\href{https://arxiv.org/abs/2109.11499}{{\ttfamily
  2109.11499}}].

\bibitem{Abe:2022mlc}
T.~Abe and Y.~Hamada, \emph{{A model of pseudo-Nambu-Goldstone dark matter from
  a softly broken $SU(2)$ global symmetry with a $U(1)$ gauge symmetry}},
  \href{https://arxiv.org/abs/2205.11919}{{\ttfamily 2205.11919}}.

\bibitem{Maleki:2022zuw}
K.~R. Maleki and K.~Ghorbani, \emph{{Loop enhancement of direct detection cross
  section in a fermionic dark matter model}},
  \href{https://arxiv.org/abs/2211.12102}{{\ttfamily 2211.12102}}.

\bibitem{Liu:2022evb}
D.-Y. Liu, C.~Cai, X.-M. Jiang, Z.-H. Yu and H.-H. Zhang, \emph{{Ultraviolet
  completion of pseudo-Nambu-Goldstone dark matter with a hidden U(1) gauge
  symmetry}}, \href{https://doi.org/10.1007/JHEP02(2023)104}{\emph{JHEP}
  {\bfseries 02} (2023) 104},
  [\href{https://arxiv.org/abs/2208.06653}{{\ttfamily 2208.06653}}].

\bibitem{Cho:2021itv}
G.-C. Cho, C.~Idegawa and E.~Senaha, \emph{{Electroweak phase transition in a
  complex singlet extension of the Standard Model with degenerate scalars}},
  \href{https://doi.org/10.1016/j.physletb.2021.136787}{\emph{Phys. Lett. B}
  {\bfseries 823} (2021) 136787},
  [\href{https://arxiv.org/abs/2105.11830}{{\ttfamily 2105.11830}}].

\bibitem{Cho:2022our}
G.-C. Cho, C.~Idegawa and E.~Senaha, \emph{{CP-violating effects on
  gravitational waves in a complex singlet extension of the Standard Model with
  degenerate scalars}},
  \href{https://doi.org/10.1103/PhysRevD.106.115012}{\emph{Phys. Rev. D}
  {\bfseries 106} (2022) 115012},
  [\href{https://arxiv.org/abs/2205.12046}{{\ttfamily 2205.12046}}].

\bibitem{Froggatt:1995rt}
C.~D. Froggatt and H.~B. Nielsen, \emph{{Standard model criticality prediction:
  Top mass 173 +- 5-GeV and Higgs mass 135 +- 9-GeV}},
  \href{https://doi.org/10.1016/0370-2693(95)01480-2}{\emph{Phys. Lett. B}
  {\bfseries 368} (1996) 96--102},
  [\href{https://arxiv.org/abs/hep-ph/9511371}{{\ttfamily hep-ph/9511371}}].

\bibitem{Kannike:2020qtw}
K.~Kannike, N.~Koivunen and M.~Raidal, \emph{{Principle of Multiple Point
  Criticality in Multi-Scalar Dark Matter Models}},
  \href{https://doi.org/10.1016/j.nuclphysb.2021.115441}{\emph{Nucl. Phys. B}
  {\bfseries 968} (2021) 115441},
  [\href{https://arxiv.org/abs/2010.09718}{{\ttfamily 2010.09718}}].

\bibitem{Cho:2022zfg}
G.-C. Cho, C.~Idegawa and R.~Sugihara, \emph{{A complex singlet extension of
  the standard model and multi-critical point principle}},
  \href{https://doi.org/10.1016/j.physletb.2023.137757}{\emph{Phys. Lett. B}
  {\bfseries 839} (2023) 137757},
  [\href{https://arxiv.org/abs/2212.13029}{{\ttfamily 2212.13029}}].

\bibitem{Azevedo:2018exj}
D.~Azevedo, M.~Duch, B.~Grzadkowski, D.~Huang, M.~Iglicki and R.~Santos,
  \emph{{One-loop contribution to dark-matter-nucleon scattering in the
  pseudo-scalar dark matter model}},
  \href{https://doi.org/10.1007/JHEP01(2019)138}{\emph{JHEP} {\bfseries 01}
  (2019) 138}, [\href{https://arxiv.org/abs/1810.06105}{{\ttfamily
  1810.06105}}].

\bibitem{Ishiwata:2018sdi}
K.~Ishiwata and T.~Toma, \emph{{Probing pseudo Nambu-Goldstone boson dark
  matter at loop level}},
  \href{https://doi.org/10.1007/JHEP12(2018)089}{\emph{JHEP} {\bfseries 12}
  (2018) 089}, [\href{https://arxiv.org/abs/1810.08139}{{\ttfamily
  1810.08139}}].

\bibitem{Alanne:2020jwx}
T.~Alanne, N.~Benincasa, M.~Heikinheimo, K.~Kannike, V.~Keus, N.~Koivunen
  et~al., \emph{{Pseudo-Goldstone dark matter: gravitational waves and
  direct-detection blind spots}},
  \href{https://doi.org/10.1007/JHEP10(2020)080}{\emph{JHEP} {\bfseries 10}
  (2020) 080}, [\href{https://arxiv.org/abs/2008.09605}{{\ttfamily
  2008.09605}}].

\bibitem{Glaus:2020ihj}
S.~Glaus, M.~M\"uhlleitner, J.~M\"uller, S.~Patel, T.~R\"omer and R.~Santos,
  \emph{{Electroweak Corrections in a Pseudo-Nambu Goldstone Dark Matter Model
  Revisited}}, \href{https://doi.org/10.1007/JHEP12(2020)034}{\emph{JHEP}
  {\bfseries 12} (2020) 034},
  [\href{https://arxiv.org/abs/2008.12985}{{\ttfamily 2008.12985}}].

\bibitem{Barger:2008jx}
V.~Barger, P.~Langacker, M.~McCaskey, M.~Ramsey-Musolf and G.~Shaughnessy,
  \emph{{Complex Singlet Extension of the Standard Model}},
  \href{https://doi.org/10.1103/PhysRevD.79.015018}{\emph{Phys. Rev. D}
  {\bfseries 79} (2009) 015018},
  [\href{https://arxiv.org/abs/0811.0393}{{\ttfamily 0811.0393}}].

\bibitem{Passarino:1978jh}
G.~Passarino and M.~J.~G. Veltman, \emph{{One Loop Corrections for e+ e-
  Annihilation Into mu+ mu- in the Weinberg Model}},
  \href{https://doi.org/10.1016/0550-3213(79)90234-7}{\emph{Nucl. Phys. B}
  {\bfseries 160} (1979) 151--207}.

\bibitem{vonBuddenbrock:2016rmr}
S.~von Buddenbrock, N.~Chakrabarty, A.~S. Cornell, D.~Kar, M.~Kumar, T.~Mandal
  et~al., \emph{{Phenomenological signatures of additional scalar bosons at the
  LHC}}, \href{https://doi.org/10.1140/epjc/s10052-016-4435-8}{\emph{Eur. Phys.
  J. C} {\bfseries 76} (2016) 580},
  [\href{https://arxiv.org/abs/1606.01674}{{\ttfamily 1606.01674}}].

\end{thebibliography}\endgroup
\bibliographystyle{JHEP}

\end{document}